\documentclass[10pt,a4paper]{IEEEtran} 

\usepackage{comment}
\usepackage{glossaries}
\usepackage{amssymb}
\usepackage{soul}
\usepackage{algorithm}
\usepackage{algpseudocode}
\usepackage{algpseudocode}
\usepackage{booktabs}
\usepackage{lettrine}
\usepackage{float}
\usepackage{mathtools}
\usepackage{placeins}
\usepackage{amsthm}
\usepackage{stfloats} 
\usepackage{colortbl} 
\usepackage{float}
\usepackage{threeparttable}
\usepackage{hyperref}
\usepackage{url}
\usepackage{pifont}
\usepackage{array}
\usepackage{tikz}

\newacronym{3d}{3-D}{three-dimensional}
\newacronym{3gpp}{3GPP}{3rd Generation Partnership Project}
\newacronym{5g}{5G}{fifth-generation}
\newacronym{6g}{6G}{sixth-generation}
\newacronym{ao}{AO}{alternating optimization}
\newacronym{aoa}{AoA}{angle of arrival}
\newacronym{aod}{AoD}{angle of departure}
\newacronym{as}{AS}{antenna selection}
\newacronym{awgn}{AWGN}{additive white gaussian noise}
\newacronym{agd}{AGD}{adaptative gradient descend}
\newacronym{bs}{BS}{base station}
\newacronym{bcrlb}{BCRLB}{bayesian Cramér-Rao lower bound}
\newacronym{b6g}{B6G}{beyond six generation}
\newacronym{ber}{BER}{bit error rate}
\newacronym{bcd}{BCD}{block coordinate descent}
\newacronym{csi}{CSI}{channel state information}
\newacronym{chest}{CHEST}{channel estimation}
\newacronym{ccdf}{CCDF}{complementary cumulative distribution function}
\newacronym{cdf}{CDF}{complementary cumulative distribution function}
\newacronym[plural=CCBs]{ccb}{CCB}{channel coherence block}
\newacronym{crlb}{CRLB}{Cramér-Rao lower bound}
\newacronym{cg}{CG}{conjugated gradient}
\newacronym{dl}{DL}{downlink}
\newacronym{doa}{DoA}{direction of arrival}
\newacronym{dof}{DoF}{degree of freedom}
\newacronym{dfmimo}{DF-MIMO}{dual-function multiple-input multiple-output}
\newacronym{dfrc}{DFRC}{dual-function radar-communication}
\newacronym{ee}{EE}{energy efficiency}
\newacronym{er}{ER}{ergodic rate}
\newacronym{epa}{EPA}{equal power allocation}
\newacronym{fim}{FIM}{Fisher information matrix}
\newacronym{fp}{FP}{fractional programming}
\newacronym{fpga}{FPGA}{field programmable gate array}
\newacronym{flop}{FLOP}{floating point}
\newacronym{iid}{i.i.d.}{independent and identically distributed}
\newacronym{isac}{ISAC}{integrated sensing and communication}
\newacronym{iot}{IoT}{internet of things}
\newacronym{jram}{JRAM}{joint reinforcement-analytical methodology}
\newacronym{kkt}{KKT}{Karush–Kuhn–Tucker}
\newacronym{ldt}{LDT}{Lagrangian Dual Transform}
\newacronym{los}{LoS}{line-of-sight}
\newacronym{mimo}{MIMO}{multiple-input multiple-output}
\newacronym{m-mimo}{mMIMO}{massive multiple-input multiple-output}
\newacronym{mrc}{MRC}{maximum ratio combining}
\newacronym{mm}{MM}{majoration-minimization}
\newacronym{mcs}{MCs}{Monte-Carlo simulation}
\newacronym{mdp}{MDP}{Markov decision process}
\newacronym{mu}{MU}{multi-user}
\newacronym{mt}{MT}{multi-target}
\newacronym{mrt}{MRT}{maximum-ratio transmission}
\newacronym{mse}{MSE}{mean-squared error}
\newacronym{mle}{MLE}{maximum likelihood estimation}
\newacronym{map}{MAP}{maximum a posteriori}
\newacronym{mo}{MO}{manifold optimization}
\newacronym{mi}{MI}{mutual information}
\newacronym{music}{MUSIC}{multiple signal classification}
\newacronym{moo}{MOO}{multi-objective optimization}
\newacronym{mui}{MUI}{multi-user interference}
\newacronym{nlos}{NLoS}{non-line-of-sight}
\newacronym{nlp}{NLP}{non-linear problem}
\newacronym{om}{OM}{oblique manifold}
\newacronym{pa}{PA}{power amplifier}
\newacronym{pdf}{PDF}{probability density function}
\newacronym{pc}{PC}{pilot contamination}
\newacronym{pin}{PIN}{positive-intrinsic-negative}
\newacronym{psc}{PSC}{phase shift control}
\newacronym{qos}{QoS}{quality of service}
\newacronym{qcqp}{QCQP}{quadratically constrained quadratic programming}
\newacronym{rf}{RF}{radio frequency}
\newacronym[plural=REs]{re}{RE}{reflective element}
\newacronym[plural=RISs]{ris}{RIS}{reconfigurable intelligent surface}
\newacronym{rl}{RL}{reinforcement learning}
\newacronym{rcs}{RCS}{radar cross section}
\newacronym{rmo}{RMO}{Riemannian Manifold Optimization}
\newacronym{rcg}{RCG}{Riemannian Conjugate Gradient}
\newacronym{rmse}{RMSE}{root mean squared error}
\newacronym{rcrlb}{RCRLB}{root Cramer-Rao lower bound}
\newacronym{se}{SE}{spectral efficiency}
\newacronym{sfp}{SFP}{sequential fractional programming}
\newacronym{sinr}{SINR}{signal-to-interference-plus-noise ratio}
\newacronym{snr}{SNR}{signal-to-noise ratio}
\newacronym{sdr}{SDR}{semidefinite relaxation}
\newacronym{sr}{SR}{sum rate}
\newacronym{sca}{SCA}{successive convex approximation}
\newacronym{sdp}{SDP}{semidefinite program}
\newacronym{soc}{SOC}{second-order cone}
\newacronym{sgcdf}{SGCDF}{Sensing-Guided Communication Dual-Function}
\newacronym{svd}{SVD}{singular value decomposition}
\newacronym{sp}{SP}{subproblem}
\newacronym{st}{ST}{single-target}
\newacronym{tdd}{TDD}{time-division duplex}
\newacronym[plural=UEs, firstplural=users' equipment (UEs)]{ue}{UE}{user's equipment}
\newacronym{ul}{UL}{uplink}
\newacronym{upa}{UPA}{uniform planar array}
\newacronym{uspa}{USPA}{uniform squared planar array}
\newacronym{ula}{ULA}{uniform linear array}
\newacronym{vr}{VR}{visibility region}
\newacronym{wsr}{WSR}{weighted sum-rate}
\newacronym{xl-mimo}{XL-MIMO}{extra-large scale massive MIMO}
\newacronym{zf}{ZF}{zero-forcing}

\renewcommand\qedsymbol{$\blacksquare$}
\newtheorem{theorem}{Theorem} 

\newtheorem{remark}{Remark}
\newtheorem{lemma}{Lemma}
\definecolor{teal}{RGB}{8, 126, 139}

\newcommand{\cmark}{\ding{51}}
\newcommand{\xmark}{\ding{55}}
\hypersetup{
 colorlinks=true, urlcolor=cyan, linkcolor=blue, 
 citecolor=red}
\newcolumntype{P}[1]{>{\centering\arraybackslash}p{#1}}

\begin{document}

\title{Dual-Function Beam Pattern Design for Multi-Target ISAC Systems: A Decoupled Approach}

\author{Wilson de Souza Junior,  Taufik Abrao, {Amine Mezghani}, and {Ekram Hossain}

\thanks{This work was supported in part by the National Council for Scientific and Technological Development (CNPq) of Brazil under Grants {314618/2023-6,  442945/2023-0}, and in part by a Discovery Grant from the Natural Sciences and Engineering Research Council of Canada (NSERC).}

\thanks{W. de Souza Jr is a PhD student and T. Abrão is an associate professor at the Electrical Engineering Department, State University of Londrina, PR, Brazil (e-mail: wilsoonjr98@gmail.com; taufik@uel.br).}
\thanks{Ekram Hossain and Amine Mezghani are with the Department of Electrical and Computer Engineering, University of Manitoba, Winnipeg, MB R3T 5V6, Canada (e-mail: ekram.hossain@umanitoba.ca and amine.mezghani@umanitoba.ca).}
}

\maketitle

\begin{abstract}

We investigate beam pattern design problem for mono-static \gls{mu} multi-point-target \gls{isac} systems, where a \gls{dfmimo} \gls{bs} performs downlink communication and radar sensing simultaneously. In \gls{isac} systems, sensing and communication inherently compete for resources. As communication demand increases, the beam pattern is reshaped, which might degrade the \gls{doa} sensing accuracy, measured in terms of \gls{mse} and lower-bounded by the \gls{crlb}. Since conventional joint formulations of the sensing-based problem often overlook this trade-off, our work addresses it by decomposing the sensing-based problem into two \glspl{sp}. This decomposition enables a more effective exploitation of the beam pattern’s physical properties, which we refer to as the \gls{sgcdf} beam pattern design. We further develop a low-complexity extension using the \gls{rmo} and convex closed-set projection. Simulation results confirm that the proposed method improves multi-target estimation accuracy, compared to traditional joint optimization strategies, by preserving the beam pattern, while the low-complexity version offers an excellent performance–complexity tradeoff, maintaining high accuracy with significantly reduced computational cost.

\end{abstract}

\begin{IEEEkeywords}
\gls{isac}, \gls{doa}, \gls{mse}, \gls{crlb}, \gls{rmo}.
\end{IEEEkeywords}

\section{Introduction}
 
The rapid evolution of wireless communication networks, coupled with the ever-increasing demand for intelligent and interconnected systems, drives the need for innovative technologies that transcend the traditional boundaries of communication. In this sense, future networks must not only provide high data throughput but also support emerging services that require real-time awareness and understanding of the surrounding environment. In this context, the upcoming \gls{6g} wireless communication systems and \gls{b6g} aim to encompass a wide range of advanced applications, such as autonomous vehicles, smart city infrastructure, industrial automation, and massive-scale \gls{iot} deployment. These applications require the simultaneous supply of ultra-reliable, high-speed, and precise environmental sensing, including object tracking and motion detection.

Motivated by these ambitious requirements, integrated sensing and communication (\gls{isac}) has  attracted significant attention in recent years. ISAC provides a promising paradigm that enables the joint use of \gls{rf} signals for both environmental sensing and data transmission concurrently. It  merges traditional communication architectures with advanced sensing capabilities, by jointly designing a \gls{dfrc} beam pattern \cite{10188491}. By unifying these functionalities within a shared hardware, \gls{isac} offers significant gains in \gls{se}, energy and spectrum saving, and sensing capabilities \cite{Liu2020}.

Recent advances in signal processing, optimization techniques, and machine learning, jointly with some new enabling techniques such as \gls{ris}, have contributed to the development of \gls{isac}-enabled systems, paving the way for the aforementioned applications. Despite this progress, the realization of practical \gls{isac} systems remains a difficult task due to challenges in terms of resource allocation. Unlike traditional communication systems, where resources are optimized solely for payload data transmission, \gls{isac} systems must simultaneously satisfy the performance requirements of both sensing and communication functionalities, which often have conflicting objectives. Key challenging problems, including beam pattern design, spectrum and interference management, security enhancement, and trade-offs between sensing accuracy and communication reliability, can be highly affected due to the conflicting nature of both functionalities.

Furthermore, resource allocation in \gls{isac} systems is challenging due to the lack of a unified performance metric capable of jointly quantifying the quality of both communication and sensing functionalities. In this sense, since the communication and sensing functionalities use distinct nature criteria, the resource allocation problem in \gls{isac} communication systems can be treated as a \textit{\gls{moo} problem} \cite{11111722}. Specifically, communication performance is typically evaluated using conventional metrics such as \gls{sr}, individual user rate, or even \gls{ber}, denoted here as $\mathcal{C}(\mathcal{R})$, whereas sensing performance is assessed through metrics such as estimation accuracy, detection probability, beam pattern gain, etc., denoted as $\mathcal{S}(\mathcal{R})$. Both of these performance metrics are functions of the allocated system resources $\mathcal{R}$, which may include transmit power, bandwidth, time slots, etc. 

To address the challenges imposed by the multi-objective nature of resource allocation in \gls{isac} systems, two widely adopted approaches. The first approach involves prioritizing one objective function while constraining the other. This leads to two different strategies, denoted in this work as \textit{sensing-based optimization} and \textit{communication-based optimization}. This formulation is beneficial in scenarios where one functionality is mission-critical, and the other must remain within acceptable bounds. The second approach involves constructing a composite objective function, denoted herein as \textit{weighted sum-based optimization}, that jointly incorporates both $\mathcal{C}(\mathcal{R})$ and $\mathcal{S}(\mathcal{R})$, enabling a more balanced optimization strategy. %These two strategies are conceptually illustrated in Fig. \ref{fig:sensingMOO}.

With the above background, the following subsection provides an overview of previous works on \gls{isac} systems. 

\begin{comment}
\begin{figure}[ht!]
    \centering
\includegraphics[width=.9\linewidth]{SensingMOO.pdf}
    \caption{Illustration of how a \gls{moo} problem can be approached over the \gls{isac} systems \cite{11111722}.} 
    \label{fig:sensingMOO}
\end{figure}
\end{comment}

\subsection{Related Works}

%Many beam pattern designs for \gls{isac} systems have recently been developed, each with distinct objectives and optimization methods.

In the literature, many beam pattern designs for \gls{isac} systems have been proposed, each with distinct objectives and optimization methods.

\subsubsection{Sensing-based optimization}  \cite{9652071} addresses a single-target scenario, minimizing the \gls{crlb} under communication rate constraint. Similarly, \cite{10955684} investigates \gls{crlb} minimization for a single point target and extends the analysis to extended targets, by proposing a low-complexity solution based on \gls{agd}. %Work \cite{10138058} tackles \gls{crlb} minimization in a \gls{ris}-assisted scenario using \gls{ao}, \gls{sdr}, and \gls{sca} to efficiently solve the non-convex problem with respect to the \gls{ris} phase shifts and \gls{dfrc} waveform. \gls{crlb} minimization is studied for a single-target \gls{ris}-assisted scenario. 
For multi-target cases, \cite{10380213} minimizes the maximum \gls{crlb} using conventional \gls{sdr} to handle the non-convex problem. Meanwhile, \cite{10938377} formulates a min-max problem minimizing the worst-case \gls{crlb} in a \gls{ris}-assisted multi-target setup, while also considering individual user rates. A product of Riemannian manifolds is used for joint optimization of \gls{ris} phase shifts and \gls{dfrc} beam pattern. For multi-user, multi-target \gls{isac} systems, \cite{10097000} investigates the \gls{crlb} minimization where a Schur complement is utilized with \gls{sdr}. The work in \cite{10756646} studies various precoding schemes and shows that sensing-based conventional communication precoding with optimized power allocation significantly reduces the computational complexity.

\subsubsection{Communication-based optimization}  %Reference \cite{10130707} focuses on \gls{sr} maximization with limited mutual interference between communication and sensing in a \gls{mu} setup for \gls{ris}-assisted single-target systems. The \gls{fp} and \gls{rmo} techniques are utilized to tackle the problem. 
\cite{11126093} aims to maximize the sum rate (\gls{sr}), while guaranteeing minimum \gls{crlb} and maximum secrecy leakage limit. The \gls{bcd} with \gls{fp}, and \gls{sca} methods are utilized to update the variables. Reference \cite{10762897} seeks to jointly maximize the system \gls{sr} and the max-min rate, employing a parallel Riemannian algorithm for efficient optimization. For multi-target scenarios, \cite{10737380} aims to maximize the communication \gls{sr} while satisfying sensing beampattern gain and users’ minimum rate constraints, solved via an \gls{rmo} and \gls{fp} approach to handle the non-convexity. Reference \cite{9591331} considers a \gls{ris}-aided \gls{mu} \gls{isac} system and aims to minimize \gls{mui} under a \gls{doa} estimation \gls{crlb} constraint, balancing interference mitigation with sensing accuracy. \cite{10411853} focuses on a \gls{ris}-assisted \gls{isac} system, aiming to minimize the transmit power while guaranteeing the individual rate and \gls{mi}.

\subsubsection{Weighted sum-based optimization}  \cite{10888351} proposed a low-complexity methodology for a single-target system.  \cite{10065868} proposes to minimize the beam pattern error, cross-correlation pattern, and the total interference of radar and multiple users for a \gls{ris}-assisted multi-user multi-target system.

Table \ref{table:ref} succinctly summarizes the above discussion and provides a clear comparison among related works in the \gls{isac} domain. It highlights the key differences in system design, problem formulation, and solution methodology.

\begin{table*}[!t] 
\centering
\caption{Comparison of Related \gls{isac} Studies}
\renewcommand{\arraystretch}{1.2}
\setlength{\tabcolsep}{4pt}
\begin{tabular}{P{1.4cm} P{3cm} P{2.6cm} P{2.3cm} P{2.2cm} P{3.7cm} P{0.5cm}}
\rowcolor[gray]{0.9}
\textbf{Ref. / Year} & \textbf{Solution-based} & \textbf{Objective} & \textbf{$\mathcal{S}(\mathcal{R})$} & \textbf{$\mathcal{C}(\mathcal{R})$} & \textbf{Scenario} & \textbf{\gls{ris}} 
\\
\toprule
\cite{9652071}, 2022 
& Schur complement, \gls{sdr} 
& Sensing-based 
& \gls{crlb} 
& Individual rate 
& \gls{dfrc} \gls{isac}, MU, single-target
&
\xmark
\\
\cite{10955684}, 2025 
& Bisection, AGD 
& Sensing-based 
& \gls{crlb} 
& Individual rate 
& \gls{dfrc} \gls{isac}, MU, single-target
& \xmark
\\
%\cite{10138058}, 2023 
%& AO, \gls{sdr}, SCA 
%& Sensing-based 
%& \gls{crlb} 
%& Individual rate 
%& \gls{dfrc} \gls{isac}, MU, single-target
%& \cmark
%\\
\cite{10380213}, 2024 
& \gls{sdr} 
& Sensing-based 
& Max. \gls{crlb} 
& Individual rate 
& \gls{dfrc} \gls{isac}, MU, multi-target 
& \xmark
\\
\cite{10938377}, 2025 
& Parallel \gls{rmo} 
& Sensing-based 
& Max. \gls{crlb} 
& Individual rate 
& \gls{dfrc} \gls{isac}, MU, multi-target
& \cmark
\\
\cite{10097000}, 2023
& Schur complement, \gls{sdr}
& Sensing-based
& Sum of \gls{crlb}
& \gls{sr}
& \gls{dfrc} \gls{isac}, \gls{mu}, multi-target
& \xmark
\\
\cite{10756646}, 2025 
& Schur complement 
& Sensing-based 
& Sum of \gls{crlb} 
& Individual rate  
& \gls{dfrc} \gls{isac}, MU, multi-target
& \xmark
\\
\bottomrule
%\cite{10130707}, 2023 
%& \gls{ao}, \gls{rmo}, \gls{fp}
%& Communication-based 
%&  Mutual Interf.
%& \gls{sr}
%& \gls{dfrc} \gls{isac}, \gls{mu}, single-target
%& \cmark 
%\\
\cite{11126093}, 2025 
& \gls{fp}, \gls{bcd}, \gls{sca} & Communication-based 
& \gls{crlb}
& \gls{sr}
& \gls{dfrc} \gls{isac}, \gls{mu}, single-target
& \cmark
\\
\cite{10762897}, 2025 
& Parallel \gls{rmo} 
& Communication-based 
& \gls{crlb} 
& \gls{sr} \& min. rate 
& \gls{dfrc} \gls{isac}, MU, single-target
& \cmark
\\
\cite{10737380}, 2025 
& \gls{rmo}, \gls{fp} 
& Communication-based & Beampattern gain
& \gls{sr} 
&  \gls{dfrc} \gls{isac} \gls{mu}, multi-target
&\xmark
\\
\cite{9591331}, 2022 
& \gls{ao}, \gls{rmo} 
& Communication-based 
& \gls{crlb}
& \gls{mui} 
& \gls{dfrc} \gls{isac}, \gls{mu}, multi-target 
& \cmark
\\
\cite{10411853}, 2024 
& \gls{ao}, \gls{sca}, \gls{rmo}
& Communication-based 
& MI
& Individual rate
& \gls{dfrc} \gls{isac}, \gls{mu}, multi-target
& \cmark
\\
\bottomrule
\cite{10888351}, 2025 
& \gls{sca} 
& Weighted sum-based 
& \gls{crlb} 
& \gls{sr}
& \gls{dfrc} \gls{isac}, \gls{mu}, single-target 
& \xmark 
\\
\cite{10065868}, 2022 
& \gls{ao}, \gls{rmo} 
& Weighted sum-based 
& Beampattern error \& cross-correlation & \gls{mui} & \gls{dfrc} \gls{isac}, \gls{mu}, multi-target, 
& \cmark 
\\
\bottomrule
\bf Proposed Method & {RMO, Convex Closed-Set Projection} & Sensing-based (decoupled) & {Sum of CRLB} & {Individual rate} & {DFRC ISAC, MU, multi-target}& \xmark \\
\bottomrule
\end{tabular}
\label{table:ref}
\footnotesize{}
\end{table*}

\subsection{Motivations and Contributions}

Designing dual-function beam patterns for \gls{isac} systems is inherently challenging, as it requires a careful balance between sensing and communication performance. The overall performance is usually characterized by sensing metrics—such as beam pattern gain, \gls{mi}, and the \gls{mse}—alongside communication metrics like the achievable rate, or the total sum-rate. Among these, specifically for sensing, the \gls{crlb} stands out in multi-target scenarios, since it establishes a fundamental lower bound on the estimation \gls{mse}, making it a powerful criterion for guiding effective beam pattern design. However, conventional approaches often adopt a joint formulation that optimizes sensing and communication objectives concurrently. Since the \gls{mse} is strongly influenced by beam pattern gain, joint dual-function \gls{crlb}-based formulations may yield inefficient beam patterns under stringent communication requirements, since they overlook the physical pattern. To address these limitations, we propose decomposing the sensing-based problem into two subproblems, which allows overcoming the inherent drawbacks of conventional joint formulations. This decoupled formulation enables us to account for two different sensing metrics: the \gls{crlb} as well as the beam pattern gain, while ensuring the communication performance.

Furthermore, unlike prior works that rely on alternative formulations to handle the \gls{crlb} in multi-target scenarios, which introduce relaxation gaps and increase computational complexity, our approach directly tackles the original problem formulation. The main contributions of this paper are threefold and summarized as follows.

\begin{itemize}
    \item  First, we consider the dual-function beam pattern design for a mono-static \gls{dfmimo} system, in which a multi-antenna \gls{bs} performs downlink multi-user communication  and \gls{doa} estimation for a multiple target scenario. The sensing-based optimization problem is  split into two sub problems. The first sub-problem is dedicated to designing an adequate beam pattern specifically for sensing by solving the \gls{crlb} minimization problem, while the second one focuses on transforming the obtained beam pattern into a dual-function beam pattern by projecting it on the communication-feasible set by minimizing the mismatch with respect to the obtained sensing beam pattern. We show that both of the subproblems can be convex in the covariance matrix domain. This strategy is referred to as the \gls{sgcdf} methodology.

    \item Building upon the \gls{sgcdf} framework, we further develop a low-complexity solution by solving the optimization problem in the beamforming domain, which reduces the problem’s dimensionality. To efficiently tackle this problem, we integrate \gls{rmo} with a projection onto a convex closed set. This combination accelerates convergence, resulting in a computationally efficient algorithm well-suited for practical implementation. Moreover, we demonstrate that the proposed low-complexity solution covers the non-dedicated sensing stream case by appropriately projecting the initial point solution. This generalization broadens the applicability to different practical system configurations.

    \item Finally, extensive simulations show that this two-step procedure better preserves the 
    inherent physical properties of the sensing-only beam pattern while meeting minimum communication \gls{qos} requirements. By decoupling sensing and communication design, it outperforms traditional joint optimization methods, achieving significantly improved \gls{doa} estimation accuracy. Additionally, the proposed low-complexity method is shown to offer an excellent performance-complexity trade-off.
        
\end{itemize}

The remainder of this paper is organized as follows. Section~\ref{sec:model} introduces the system model. Section~\ref{section:problem} presents the problem formulation, including the system assumptions, objectives, and constraints that guide the proposed approach. In Section~\ref{sec:propSol}, we describe the proposed solution to the formulated problem. Simulation results and a comprehensive performance analysis are presented in Section~\ref{sec:results}. Finally, Section~\ref{sec:conclusion} concludes the paper and outlines potential directions for future research.

\section{System Model and Assumptions}\label{sec:model}

We consider a mono-static \gls{isac} system with a coherence interval of $L$ transmission blocks. In such a system, a \gls{bs} with co-located $M_T$ transmit antennas and $M_R$ receive antennas serves $K$ communication \glspl{ue} while simultaneously detecting $T$ targets. The dual function MIMO (DF-MIMO) \gls{bs} is assumed to operate in full-duplex mode with perfect self-interference mitigation \cite{10762897}. The transmitted signal $\boldsymbol{x}_{\ell} \in \mathbb{C}^{M_T}$ at time slot $\ell$ can be expressed as
\begin{equation}
    \boldsymbol{x}_{\ell} = \boldsymbol{W}_{C}\boldsymbol{c}_{\ell}+ \boldsymbol{W}_{S}\boldsymbol{s}_{\ell} = \boldsymbol{W}\boldsymbol{\tilde{x}}_\ell,
\end{equation}
where $\boldsymbol{c}_\ell \triangleq [c_{1,\ell},...,c_{K,\ell}]^T$ is the communication symbols intended to the communication \glspl{ue} during the $\ell$-th transmission block, and $\boldsymbol{s}_{\ell}$ is the probing signal devoted to radar sensing. Finally, $\boldsymbol{W}_{C} = [\boldsymbol{w}_{c,1}, \boldsymbol{w}_{c,2},\dots,\boldsymbol{w}_{c,K}] \in \mathbb{C}^{M_{T} \times K}$ is the communication beamforming matrix applied to $\boldsymbol{c}_{\ell}$, and $\boldsymbol{W}_{S} = [\boldsymbol{w}_{s,1}, \boldsymbol{w}_{s,2},\dots,\boldsymbol{w}_{s,N_t}] \in \mathbb{C}^{M_T \times M_{T}}$ is the sensing beamforming matrix, applied to $\boldsymbol{s}_{\ell}$. Let us define the joint dual-function beamforming augmented matrix $\boldsymbol{W} = \begin{bmatrix} \boldsymbol{W}_C, \boldsymbol{W}_S \end{bmatrix} \in \mathbb{C}^{M_T \times (K+M_T)},$ and the joint
 dual-function data augmented matrix $\mathbf{\tilde{X}} = \begin{bmatrix} \mathbf{C}, \mathbf{S} \end{bmatrix}^T=[\boldsymbol{\tilde{x}}_1,\boldsymbol{\tilde{x}}_2,\dots,\boldsymbol{\tilde{x}}_L] \in \mathbb{C}^{(K+M_T) \times L}$.

 \begin{figure}[ht!]
    \centering
    \includegraphics[width=0.9\linewidth]{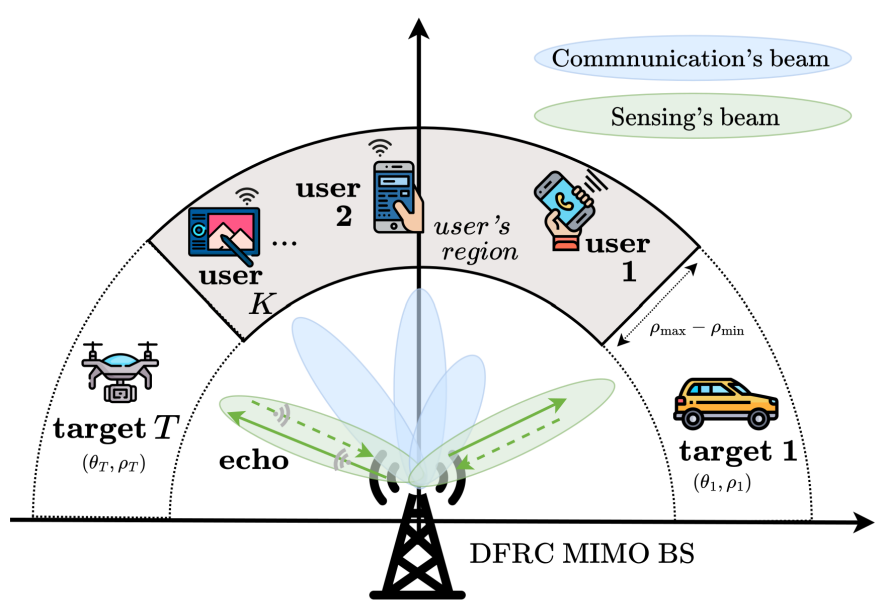}
    \caption{Illustration of the \gls{isac} system considered in this work.} 
    \label{fig:system}
\end{figure}

 When $L$ is sufficiently large, $L\rightarrow \infty$, the communication data symbols can be assumed to be independent, satisfying $\mathbb{E}[\boldsymbol{C} \boldsymbol{C}^H] = \frac{1}{L} \boldsymbol{C}\boldsymbol{C}^H  = \mathbf{I}_{K}$. At the same time, the sensing signal is also carefully designed to satisfy $\frac{1}{L}\boldsymbol{S}\boldsymbol{S}^H = \mathbf{I}_{M_T}$. We assume that there is no correlation between sensing data and communication data, which means that $K+M_T$ data streams are uncorrelated, satisfying 
 \begin{equation}
     \mathbb{E}[\boldsymbol{\tilde{X}}\boldsymbol{\tilde{X}}^H] =\frac{1}{L}\boldsymbol{\tilde{X}}\boldsymbol{\tilde{X}}^H = \mathbf{I}_{K+M_T}.
 \end{equation}
Therefore, the sample covariance matrix of beam pattern $\boldsymbol{X}=[\boldsymbol{x}_1,\boldsymbol{x}_2, \dots,\boldsymbol{x}_L]$ can be expressed as
\begin{align} 
    \boldsymbol{R}_X = \frac{1}{L} \boldsymbol{X}\boldsymbol{X}^H  
    = \boldsymbol{W}\boldsymbol{W}^H 
    %&= \boldsymbol{W_C}\boldsymbol{W_C}^H + \boldsymbol{W_S}\boldsymbol{W_S}^H\nonumber \\
    = \boldsymbol{R}_C + \boldsymbol{R}_S.
    \label{eq:rx}
\end{align}

It is noteworthy that, in the adopted scheme, both communication signals and sensing beam patterns are precoded by $\boldsymbol{W}$, meaning that the covariance matrix has as many as possible \gls{dof}, i.e., the covariance matrix $\boldsymbol{R}_X$ is full-rank, given as $M_t$ \cite{9652071}.

\subsection{Communication Model}

It is assumed that each \gls{ue} is equipped with a single antenna. The signal vector received by the $k$-th \gls{ue} is given by
\begin{equation}
    \boldsymbol{y}_k =  \boldsymbol{X}^H\boldsymbol{h}_{k} + \boldsymbol{n}_k,
    \label{eq:1}
\end{equation}
where $\boldsymbol{n}_k$ represents \gls{awgn}, $\boldsymbol{n}_k \sim \mathcal{CN}(\mathbf{0},\sigma^2\mathbf{I}_{L}) ~\forall k$, while $\boldsymbol{h}_k$ denotes the channel vector corresponding to \gls{ue} $k$. The \gls{sinr} of \gls{ue} $k$ can be written as
\begin{equation}
    \operatorname{SINR}_k = \frac{\left| \boldsymbol{h}_k^H \boldsymbol{w}_{c,k} \right|^2}{\sum_{j=1, j\neq k}^{K} \left| \boldsymbol{h}_k^H \boldsymbol{w}_{c,j} \right|^2 + \sum_{t=1}^{M_T} | \boldsymbol{h}_k^H\boldsymbol{w}_{s,t}|^2 + \sigma_k^2},
    \label{eq:sinr}
\end{equation}
where the term $\sum_{i=1,i\neq k}^{K} \left| \boldsymbol{h}_k^H \boldsymbol{w}_{c,i} \right|^2$ is the multi-user interference and the term $\sum_{t=1}^{M_T} | \boldsymbol{h}_k^H\boldsymbol{w}_{s,t}|^2$ denotes the interference caused by the dedicated sensing signal to the communication. The normalized rate of the $k$-th \gls{ue} can be given as
\begin{equation}\label{eq:rate}
    R_k =  \log_2(1+\operatorname{SINR}_{k}).
\end{equation}
The communication rate serves as a key metric for  communication performance. 

We assume a Rician fading environment for the communication users' channel links. The channel matrix $\boldsymbol{H} = [\boldsymbol{h}_{1}, \boldsymbol{h}_{2},\dots,\boldsymbol{h}_{K} ]$, which is assumed to be perfectly known\footnote{Perfect \gls{csi} is assumed to evaluate the best achievable performance of the multi-target \gls{isac} system. The proposed scheme can also be adapted to imperfect \gls{csi} scenarios, where beamforming is optimized to maximize the ergodic rate based on statistical \gls{csi}, which is more robust to estimation errors than instantaneous \gls{csi}.}, is given as 
\begin{equation}
    \boldsymbol{h}_{k} = \sqrt{\frac{\beta_{k}\kappa}{\kappa + 1}}\boldsymbol{h}_{k}^{{\rm LoS}} + \sqrt{\frac{\beta_{k}}{\kappa + 1}} \boldsymbol{h}_{2,k}^{{\rm NLoS}},
\end{equation}
where $\kappa$ is the Rician $K$-factor, and $\beta_{k}$ is the path-loss for \gls{bs}-\glspl{ue} link. Besides, $\boldsymbol{h}_k^{\rm LoS} = \boldsymbol{a}_T(\theta^{\operatorname{UE}}_k)$ represents the channels associated with the \gls{los} component of $k$-th \gls{ue}, while $\boldsymbol{h}_k^{\rm NLoS}$ represents the channels associated with the \gls{nlos} component, whose entries are \gls{iid} and follow the complex Gaussian distribution with zero mean and unit variance.

\subsection{Radar Model}

The \gls{dfmimo} \gls{bs} sends a sequence of probing signals, which will be reflected by the target back to the receiver antennas at the \gls{dfmimo} \gls{bs}; the unknown \gls{doa} is then estimated by processing the received echo signals. The collection of received signals over $L$ samples at the \gls{dfmimo} \gls{bs} is thus given by
\begin{equation} \label{eq:Y}
    \boldsymbol{\tilde{Y}}=[\boldsymbol{\tilde{y}}_1,\dots,\boldsymbol{\tilde{y}}_L] = \left(\sum_{t=1}^{T}\alpha_t \boldsymbol{G}(\theta_t)\right)\boldsymbol{X} + \boldsymbol{N},
\end{equation}
where $\boldsymbol{N} = [\boldsymbol{n}_1,\dots,\boldsymbol{n}_L]$ denotes the \gls{awgn} and $\alpha_t \in \mathbb{C}$ is the complex-valued \gls{rcs}, which also includes the effect of round-trip path loss coefficient.

We consider a \gls{los} propagation environment where no obstruction/scatter exists between the \gls{dfmimo} \gls{bs} and each possible target location. The overall channel from the \gls{dfmimo} transmitter to the \gls{dfmimo} receiver via target reflection is given by 
\begin{equation}
    \boldsymbol{G}(\theta_t) = 
    \boldsymbol{a}_R(\theta_t) \boldsymbol{a}_T^H(\theta_t), \quad \forall t
\end{equation}
where $\boldsymbol{a}_T(\cdot)$ and $\boldsymbol{a}_R(\cdot)$ are the transmit and receive steering vectors, respectively:
\begin{equation}
    \boldsymbol{a}(\theta) = [1, e^{j\pi \sin(\theta)}, \dots e^{j\pi (N-1) \sin(\theta)}]^T.
\end{equation}

From a radar perspective, the goal is to estimate the parameters after $L$ transmission blocks, where the performance is measured in terms of the \gls{mse}:
\begin{equation}
    \operatorname{MSE} = \mathbb{E}\left[ \|\boldsymbol{\zeta} - \boldsymbol{\hat{\zeta}}\|^2 \right],
\end{equation}
where $\boldsymbol{\zeta}$ denotes the vector of unknown parameters to be estimated. This paper focuses thoroughly on the estimation of target's azimuth angles $\theta_t$, since they are embedded in a non-linear structure involving the array response. This is different from the \gls{rcs} $\alpha_t$, which can be estimated through straightforward techniques, such as \gls{mle}, once the azimuth angles are estimated \cite{9500663,Song2023}, \cite{10938377}. Therefore, the unknown parameters vector, which remains constant over the coherence interval, is defined as
$\boldsymbol{\zeta}=[\theta_1, \theta_2, \dots, \theta_T]^T \in \mathbb{R}^{T\times 1}$. 

Since a closed-form expression for \gls{mse} is hard to obtain, the \gls{crlb} is often assumed to be an optimization criterion, since it is a lower-bound for the variance of an unknown estimated with any unbiased estimator \cite{Kay97}:  
\begin{equation}
    \operatorname{Var}(\hat{\theta}_t) \geq \operatorname{CRLB}(\theta_t) = \Big[\boldsymbol{F}^{-1}\Big]_{t,t},
\end{equation}
with $\operatorname{Var}(\cdot)$ being the variance operator and $\boldsymbol{F}$ being the Fisher information matrix, where the element in the $i$-th row and $j$-th column is expressed as \cite{1703855}
\begin{equation} \label{eq:Fisherik}
    [\boldsymbol{F}]_{i,j} = \frac{2L}{\sigma^2}\Re\left\{ \alpha_i^* \alpha_j \operatorname{tr}\left( \dot{\boldsymbol{G}}(\theta_j)\boldsymbol{R}_X \dot{\boldsymbol{G}}^H(\theta_i) \right)\right\},
\end{equation}
where $\dot{\boldsymbol{G}}(\theta_t)=\frac{\partial\boldsymbol{G}(\theta_t)}{\partial \theta_t}$ is the partial derivative of $\boldsymbol{G}(\theta_t)$ w.r.t. $\theta_t$. Besides, $\boldsymbol{F}$ is a real symmetric matrix $\boldsymbol{F} \in \mathbb{R}^{T\times T}$.

One can see that the performance of estimating $\theta_t$ is critically dependent on the dual function beam pattern $\boldsymbol{X}$, which is, by itself, essentially related to the design of the dual function beamforming matrix $\boldsymbol{W}$. In the following, we will investigate the optimization of the \gls{dfmimo} beamforming matrix.

\section{Problem Formulation and Classical Solution}\label{section:problem}

\subsection{Problem Formulation}
We consider an \gls{isac} system, where the sensing task is integrated into the conventional communication system. Accordingly, the considered system is designed to optimize the covariance matrix of the dual-function transmit signal, $\boldsymbol{R}_X$, to guarantee \gls{qos} requirements for each \gls{ue} in terms of rate, while simultaneously minimizing the overall estimation error of the \gls{doa} associated with each target. To this end, we employ the \textit{Trace-Opt} criterion \cite{4359542}, which represents the sum of each of the \gls{crlb} values regarding the \gls{doa} estimation. The joint sensing-communication problem is formulated as follows:
\begin{subequations} \label{eq:problem1}
 \begin{align} 
    \underset{\boldsymbol{W}}{\operatorname{minimize}} & \quad \operatorname{tr}\left( \boldsymbol{F}^{-1}\right)
     \\
     \operatorname{subject~to}  
     & ~ \left[ \boldsymbol{W}\boldsymbol{W}^H\right]_{m,m} \leq \frac{P_{\max}}{M_T} \quad \forall m \label{eq:c1}
      \\
     & ~ \Gamma_k |\boldsymbol{w}_k^H\boldsymbol{h}_k|^2  \geq  \sum_{j=1}^{K+M_T}|\boldsymbol{w}_j^H\boldsymbol{h}_k|^2 + \sigma^2, ~ \forall k
     \label{eq:c2}
 \end{align}
\end{subequations}
where  $P_{\max}$ denotes the total transmit power budget, $\Gamma_k = \left(1+\gamma_k^{-1}\right)$, with $\gamma_k=2^{R_{\min,k}}-1$, and $R_{\min,k}$ is the minimum rate for the $k$-th \gls{ue}. Constraint \eqref{eq:c1} limits the consumed power per antenna, and constraint \eqref{eq:c2} assures a minimum rate for each communication user.

The presence of the matrix inversion in the objective function, combined with the non-convexity of constraint \eqref{eq:c2}, makes problem \eqref{eq:problem1}  non-convex and challenging to solve. In what follows, we first revisit the classical \gls{sdr} solution, and then propose a new method to solve \eqref{eq:problem1}.

\subsection{Classical SDR Solution} \label{ss:SDR}

By denoting $\boldsymbol{R}_{C,k}=\boldsymbol{w}_{c,k}\boldsymbol{w}_{c,k}^H~\forall k$, and $\boldsymbol{R}_C = \sum_{k=1}^{K} \boldsymbol{R}_{C,k}$, problem \eqref{eq:problem1} can be lifted to the covariance matrices domain, and cast as the following \gls{sdp}: 
\begin{subequations} \label{eq:problemSDP}
 \begin{align} 
     \underset{\{\boldsymbol{R}_{C,k}\}, \boldsymbol{R}_X}{\operatorname{minimize}} & ~ \operatorname{tr}\left( \boldsymbol{F}^{-1}\right)
     \\  \operatorname{subject~to}
     & ~ \left[ \boldsymbol{R}_X\right]_{m,m} \leq \frac{P_{\max}}{M_T} \quad \forall m \label{eq:c11}
     \\
      & ~ \hspace{-8mm}\Gamma_k\operatorname{tr}\left( \boldsymbol{R}_{C,k}\boldsymbol{h}_k\boldsymbol{h}_k^H \right) \geq  \operatorname{tr}\left(\boldsymbol{R}_X \boldsymbol{h}_k\boldsymbol{h}_k^H \right) + \sigma^2 ~ \forall k
      \label{eq:c22}
      \\
      & ~   \boldsymbol{R}_{C,k} \in \mathcal{S}^{M_T}_+, ~\forall k ~\boldsymbol{R}_X- \boldsymbol{R}_{C} \in \mathcal{S}^{M_T}_+ \label{eq:c3}
      \\
      & ~  \operatorname{rank}(\boldsymbol{R}_{C,k}) = 1, ~ \forall k. \label{eq:c4}
 \end{align}
\end{subequations}

Constraint \eqref{eq:c3} ensures the feasibility of $\boldsymbol{R}_X$, $\boldsymbol{R}_S$, and $\boldsymbol{R}_C$, and constraint \eqref{eq:c4} is the non-convex unity rank constraint. Due to the construction of its objective function as well as the unity rank constraint \eqref{eq:c4}, \eqref{eq:problemSDP} is a non-convex problem and challenging to solve. A reformulation will be required to to make the problem convex. First, we consider the \gls{sdr} of the problem by removing \eqref{eq:c4}. To tackle the objective function, we should notice that the function $\operatorname{tr}(\boldsymbol{X}^{-1})$ is matrix decreasing on the positive definite matrix space \cite{boyd2004convex,10050406}; thus, by defining $\boldsymbol{U}$, with $\boldsymbol{F} \succ \boldsymbol{U}$, we have that $\operatorname{tr}(\boldsymbol{F}^{-1}) < \operatorname{tr}(\boldsymbol{U}^{-1}) $. Therefore, the relaxed problem can be equivalently expressed as follows: 
\begin{subequations}\label{eq:problemSDPrelax}
 \begin{align}
     \underset{\{\boldsymbol{R}_{C,k}\},\boldsymbol{R}_X,\boldsymbol{U}}{\operatorname{minimize}} & \quad \operatorname{tr}\left( \boldsymbol{U}^{-1}\right)
     \\
     & ~ \eqref{eq:c11}, \eqref{eq:c22}, \eqref{eq:c3}, \nonumber
     \\
     & ~ \boldsymbol{F} - \boldsymbol{U} \succ \boldsymbol{0} \label{eq:c5}.
 \end{align}
\end{subequations}

Problem \eqref{eq:problemSDPrelax} is a convex problem and can readily be solved by any convex solver. Due to the \gls{sdr}, the solution of \eqref{eq:problemSDPrelax}, $\{\boldsymbol{R}^{\star}_{C,k}\}$ might not be rank-1. When the obtained solution is rank-1, the optimal solution is obtained; otherwise, we can extract a feasible solution for \eqref{eq:problemSDPrelax}, denoted as $\boldsymbol{\tilde{w}}_{c,k}$, from the obtained $\boldsymbol{R}^{\star}_{C,k}$ as follows \cite{9124713}:
\begin{subequations} \label{eq:Rankprojec}
\begin{align}
    \boldsymbol{\tilde{w}}^{\star}_{c,k} &= (\boldsymbol{h}_k^H \boldsymbol{R}^{\star}_{C,k} \boldsymbol{h}_k)^{-1/2} \boldsymbol{R}^{\star}_{C,k} \boldsymbol{h}_{k}, ~~ \forall k
    \\
    \boldsymbol{\tilde{R}}^\star_{C,k} &= \boldsymbol{\tilde{w}}^\star_{c,k}{\boldsymbol{\tilde{w}}^{\star H}_{c,k}}, \quad \forall k
    \\
    \boldsymbol{R}^\star_S &= \boldsymbol{R}^\star_X - \sum_{k=1}^{K}\boldsymbol{\tilde{R}}^\star_{C,k}.
\end{align}
\end{subequations}

Although the classical \gls{sdr} methodology can provide a feasible solution for \eqref{eq:problem1}, we identify two major problems with respect to this solution. Both limitations are discussed in the following:
\begin{enumerate}
    \item \textit{High complexity:} The domain transformation from vectors $\{\boldsymbol{w}_{c,k}\} \in \mathbb{C}^{M_T}$ to covariance matrices $\{\boldsymbol{R}_{C,k}\} \in \mathbb{C}^{M_T\times M_T}$ introduces optimization variables of higher dimensions, which intuitively results in prohibitively high computational complexity; 

    \item \textit{Inefficient beam pattern:} Since the objective of the formulated problem is to minimize the overall \gls{crlb} associated with the estimation of \gls{doa}, while ensuring a minimum rate requirement for communication users,  it does not explicitly account for the physical irradiated power toward the targets. This limitation may,   \textit{mainly under high communication rates}, lead to  undesired power redistribution, weakening the main lobes in the direction of targets and jeopardizing the estimation accuracy.
\end{enumerate} 

Motivated by the aforementioned limitations, the next section proposes a new formulation for the sensing-based problem and discusses its solution.

\section{Revised Formulation and Proposed Solution} \label{sec:propSol}

In order to circumvent the inherent limitations of the \gls{sdr} solution, in this section, we introduce our proposed solution for problem \eqref{eq:problem1}. Firstly, we introduce a discussion about the general idea, which we term as \gls{sgcdf} beam pattern, and then formulate the associated problems. Subsequently, a low-complexity algorithm is proposed. 

\subsection{SGCD Beam Pattern Design}

We adopt a structured approach that decouples the original problem \eqref{eq:problem1} into two sequential sub problems (\gls{sp}). The central idea is to formulate a sensing-based optimization problem, denoted as $\mathcal{SP}_I$, and optimize the beam pattern exclusively for sensing performance, while ignoring communication constraints. This sensing-based optimization yields an adequate beam pattern only for sensing and is utilized in the following as the sensing-reference beam pattern. In the second stage, we construct a communication-feasible beam pattern by projecting the previously obtained solution onto the set of beam pattern that satisfy the communication requirements, which is achieved by solving $\mathcal{SP}_{II}$. This projection step aims to preserve the sensing characteristics as much as possible, seeking to mitigate the \textit{inefficient beam pattern} limitation while ensuring communication feasibility. This is different from the formulated approach in Eq. \eqref{eq:problem1}, which solves the joint problem simultaneously and aims to minimize the \gls{crlb} value, sacrificing the beam pattern.
%, which is essential for detection. 
We refer to this strategy as the \gls{sgcdf} strategy.

\subsubsection{Subproblem I}

The first \gls{sp} focuses on designing a beam pattern that is well-suited for sensing functionality. To this end, we adopt a \gls{crlb} minimization framework, formulated without considering the communication constraints, as previously discussed. The choice of the \gls{crlb} is motivated by its practical relevance in target tracking scenarios, where it can be evaluated using coarse estimates of the target parameters. Additionally, the resulting beam pattern from this \gls{sp} ensures sufficient beam pattern gain for each target. Inspired by the discussion in Section~\ref{ss:SDR}, the first \gls{sp} can be formulated as the following convex optimization problem:
\begin{subequations}\label{eq:SP1}
 \begin{align}
     \mathcal{SP}_I:\quad  \underset{\boldsymbol{R}_X,\boldsymbol{U}}{\operatorname{minimize}} & \quad \operatorname{tr}\left( \boldsymbol{U}^{-1}\right)
     \\
     \operatorname{subject~to} & \quad \eqref{eq:c11},  \eqref{eq:c3}, \eqref{eq:c5}.
 \end{align}
\end{subequations}

$\mathcal{SP}_I$ can be efficiently solved with any standard convex optimization solver. Let $\boldsymbol{R}^{\star}_{X}$ be the optimal solution of \eqref{eq:SP1}. In the next step, we formulate the second \gls{sp} in order to address the so-called \textit{inefficient beam pattern} limitation, which may arise when the joint beam pattern is altered to accommodate communication constraints, by solving problem \eqref{eq:problem1}.

\subsubsection{Subproblem II}

Specifically, the second \gls{sp} is formulated to preserve the spatial characteristics, in particular, the main lobes' power distribution of the reference beam pattern $\boldsymbol{R}^{\star}_X$, while integrating the communication requirements. Therefore, we formulate the beam pattern error problem as the following \gls{mse}-based problem:
\begin{subequations} \label{eq:SP2}
    \begin{align}
     \mathcal{SP}_{II}: \quad  \underset{\{\boldsymbol{R}_{C,k}\},\boldsymbol{\tilde{R}}_X}{\operatorname{minimize}} & \quad \| \boldsymbol{R}^{\star}_X - \boldsymbol{\tilde{R}}_X \|_{F}^2 
     \\
     \operatorname{subject~to} & \quad \eqref{eq:c11},  \eqref{eq:c22}, \eqref{eq:c3}.
    \end{align}
\end{subequations}

Problem \eqref{eq:SP2} is also a convex problem, and can be solved with, e.g., CVX. If the optimal solution $\boldsymbol{\tilde{R}}_{C,k}^{\star}$ is rank-1, then it is the optimal solution for the problem; otherwise, \eqref{eq:Rankprojec} should be utilized. The \gls{sgcdf} algorithm is summarized in \textbf{Algorithm \ref{alg:SGC}}.

\begin{algorithm}[!ht]
\caption{\hspace{-.1cm}{\bf :}  Proposed \gls{sgcdf}}\label{alg:MO-SGC-DF}
\begin{algorithmic}[1] 
\State {\bf Input:} $\boldsymbol{H}_k~\forall k$, $R_{\min}$, $\sigma^2$, $\epsilon$
\Statex \textit{Subproblem I: optimal sensing beam pattern}
\State Obtain $\boldsymbol{R}_X^{\star}$ by solving $\mathcal{SP}_I$ \eqref{eq:SP1} with convex solvers

\Statex \textit{Subproblem II: communication-feasible beam pattern}

\State Obtain  $\boldsymbol{\tilde{R}}_X^{\star}$, and $\{\boldsymbol{R}^{\star}_{C,k} \} ~\forall k$ by solving $\mathcal{SP}_{II}$ \eqref{eq:SP2} with convex solvers

\For{$k=1:K$}
    \If{$\operatorname{rank}(\boldsymbol{R}^{\star}_{C,k}) = 1$}
        \State $\boldsymbol{\tilde{R}}^{\star}_{C,k} = \boldsymbol{R}_{C,k}^{\star}$
    \Else
        \State Compute the $\boldsymbol{\tilde{w}}^{\star}_{c,k}$ and $\boldsymbol{\tilde{R}}^{\star}_{C,k}$ as \eqref{eq:Rankprojec}
    \EndIf
\EndFor 

\State Compute $\boldsymbol{R}^{\star}_S$ as \eqref{eq:Rankprojec}

\State Compute the SVD of $\boldsymbol{R}_S^{ \star} = \boldsymbol{S}\boldsymbol{\Lambda}\boldsymbol{S}^H$
\State Compute $\boldsymbol{W}_S^{\star} = \boldsymbol{S}\boldsymbol{\Lambda}^{1/2}$

\State {\bf Output:} DF-Beamforming $\boldsymbol{W}^{\star}$ matrix
\end{algorithmic}
\label{alg:SGC}
\end{algorithm}

The proposed \gls{sgcdf} algorithm can provide a feasible solution; however, since the problem is solved in the covariance matrix domain, it can also suffer from high complexity. In order to reduce the complexity, in the following sub-section, we provide a feasible solution with low complexity.

\subsection{Low-Complexity SGCDF Beam Pattern}

In this section, we propose a low-complexity optimization framework to further reduce the computational burden of the original formulation of the proposed \gls{sgcdf}. To this end, the \gls{sgcdf} approach is addressed directly in the beamforming domain, rather than in the covariance matrix domain. Leveraging the Riemannian manifold optimization (\gls{rmo}) approach, we begin by formulating and solving $\mathcal{SP}_I$, Eq. \eqref{eq:SP1}, as an unconstrained optimization problem. For $\mathcal{SP}_{II}$, \eqref{eq:SP2}, based on the obtained beam pattern in $\mathcal{SP}_{I}$, we incorporate a \gls{soc}-projection within the \gls{rmo} framework to enforce the communication-feasibility. The formulation and solution strategies for both \glspl{sp} are revisited in what follows, under the lens of the \gls{rmo}-based approach.

\subsubsection{Subproblem I}
Following the formulation provided by Eq. \eqref{eq:problem1}, the \gls{crlb} minimization problem for the beamforming domain can be expressed as 
\begin{subequations}\label{eq:SPLC1}
 \begin{align}
     \underset{\boldsymbol{W}}{\operatorname{minimize}} & \quad f_1(\boldsymbol{W})\triangleq \operatorname{tr}\left( \boldsymbol{F}^{-1}\right)
     \\
     \operatorname{subject~to} & \quad \eqref{eq:c1}.
 \end{align}
\end{subequations}

Firstly, we should notice that the objective function of \eqref{eq:SPLC1} is non-convex. Besides, it is intuitive that the transmit power should be fully utilized to enhance both functionalities' performance \cite{10938377,10762897}. To this end, we assume that each transmit antenna operates in the full power mode; thus, the constraint \eqref{eq:c1} can be written as
\begin{equation} \label{eq:oblique}
    \left[ \boldsymbol{W}\boldsymbol{W}^H \right]_{m,m} = \| [\boldsymbol{W}]_{m,:} \|^2 = \frac{P_{\max}}{M_T}, \quad \forall m.
\end{equation}

To address the challenging problem \eqref{eq:SPLC1}, we explore the fact that \eqref{eq:oblique} forms a smooth manifold in the Euclidean space, also known as complex $\mathcal{OB}$ \cite{8288677,boumal2023intromanifolds}, and defined as 
\begin{equation}
    \mathcal{OB} = \left \{ \boldsymbol{W} \in \mathbb{C}^{M_T \times (K+M_T)} \big| ~ \|\left[ \boldsymbol{W} \right]_{m,:}\| = \rho, ~\forall m\right\}.
\end{equation}

The $\mathcal{OB}(M_T,K+M_T)$ defines the set of all $M_T \times (K+M_T)$ complex matrices with $\rho=\sqrt{P_{\max}/M_T}$ row norm, i.e., each row of $\boldsymbol{W}$ lies on a complex hyper-sphere of radius $\rho$.
This geometric structure allows us to reformulate the problem \eqref{eq:SPLC1} as a constrained problem in Euclidean space to an unconstrained one over the manifold $\mathcal{OB}(M_T,K+M_T)$:
\begin{equation}\label{eq:SGCDF-RMO1}
    \mathcal{SP}^{LC}_{I}: \quad \underset{\boldsymbol{W} \in\mathcal{OB}}{\operatorname{minimize}}  \quad f_1(\boldsymbol{W}).
\end{equation} 

To tackle problem \eqref{eq:SGCDF-RMO1} with reduced computational complexity, we propose employing a gradient-descent-based algorithm.
Analogous to classical gradient descent in Euclidean spaces, this approach involves two fundamental operations: determining a descent direction that respects the manifold’s geometry and selecting a suitable step size along that direction. These operations are iteratively applied until the convergence criterion is met.

In the Riemannian setting, the manifold can be locally linearized around any point $\boldsymbol{W} \in \mathcal{OB}$ via the tangent space $\mathcal{T}_{\boldsymbol{W}}$, defined as \cite[Eq. 3.21]{10.5555/1557548}:
\begin{equation}
    \mathcal{T}_{\boldsymbol{W}}(\boldsymbol{X}) = \left\{ \boldsymbol{X} \in \mathbb{C}^{M_T\times (K+M_T)} \Big| ~ \left<\boldsymbol{W,\boldsymbol{X}}\right> = 0 \right \},
\end{equation}
where $\left<\cdot,\cdot\right>$ is a Riemannian metric defined as the real inner product in the linear space of matrices \cite[Eq. 3.14, Eq. 3.16]{10.5555/1557548}:
\begin{equation}
    \left<\boldsymbol{W},\boldsymbol{X}\right> = \Re\left\{ \operatorname{tr}\left(\boldsymbol{W} \boldsymbol{X}^H\right) \right\}.
\end{equation}

Within this framework, the Riemannian gradient $\operatorname{grad}f_1(\boldsymbol{W})$, is obtained by projecting the Euclidean gradient, $\nabla f_1(\boldsymbol{W})$ onto the tangent plane $\mathcal{T}_{\boldsymbol{W}}$, using the operator $\Pi_{\mathcal{T}_{\boldsymbol{W}}}(\cdot)$, expressed as follows:
\begin{equation}\label{eq:proj_tangent}
    \Pi_{\mathcal{T}_{\boldsymbol{W}}}(\boldsymbol{X}) = \boldsymbol{X} - \frac{1}{\rho^2} \Re\left\{ \left( \boldsymbol{W}\boldsymbol{X}^H \right) \odot \mathbf{I}_{M_T} \right\} \boldsymbol{W}.
\end{equation}
Therefore, the Riemannian gradient can be computed as follows:
\begin{align} \label{eq:Rgrad1}
    \operatorname{grad} f_1(\boldsymbol{W}) & = \Pi_{\mathcal{T}_{\boldsymbol{W}}}(\nabla f_1(\boldsymbol{W})),
    \\ 
    &=\nabla f_1(\boldsymbol{W}) - \frac{1}{\rho^2}\Re \Big\{  \nabla f_1(\boldsymbol{W}) \boldsymbol{W}^H \odot \mathbf{I}_{M_T}\Big\} \boldsymbol{W}. \nonumber
\end{align}

Besides, the Euclidean gradient of $f_1(\boldsymbol{W})$ in \eqref{eq:SPLC1} can be computed as follows:
\begin{equation} \label{eq:Egrad1}
    \nabla f_1(\boldsymbol{W}) = 2 \boldsymbol{\Omega} \boldsymbol{W}, 
\end{equation}
with $\boldsymbol{\Omega}$ being an auxiliary matrix computed as follows: 
\begin{equation} \label{eq:omega}
    \boldsymbol{\Omega} = \sum_{t=1}^{T} [\boldsymbol{F}^{-1}]_{t,t} \left( \sum_{i\neq t,j\neq t}^{T} [\boldsymbol{F}_t^{-1}]_{i,j} \boldsymbol{A}_{j,i} - \sum_{i,j}^{T} [\boldsymbol{F}^{-1}]_{i,j} \boldsymbol{A}_{j,i} \right),
\end{equation}
with $\boldsymbol{A}_{i,j} \triangleq \frac{2L}{\sigma^2} \alpha^*_i \alpha_j \dot{\boldsymbol{G}}^H(\theta_i)\dot{\boldsymbol{G}}(\theta_j)$, and $\boldsymbol{F}_t$ being identical to $\boldsymbol{F}$ except its $t$-th column has been replaced by $\mathbf{e}_t$, with $\mathbf{e}_t$ denoting the $t$-th column vector of identity matrix. %The derivation of \eqref{eq:proj_tangent} and \eqref{eq:Egrad1} is detailed in Appendix \ref{app:tangent_plane} and Appendix \ref{app:Egrad}, respectively.

Equipped with the Riemannian gradient obtained via projection mapping, we can deploy a steepest-descent procedure to find a locally optimal solution for \eqref{eq:SGCDF-RMO1}. However, the steepest descent algorithm is known to exhibit slow convergence in practice. Hence, to accelerate convergence \gls{cg} algorithm is adopted. The Riemannian generalization of the \gls{cg}, denoted as \gls{rcg}, updates the descent direction $\boldsymbol{D}$ at the $\ell$-th iteration as
\begin{equation} \label{eq:descent_direction}
    \boldsymbol{D}_{\ell} = 
    \begin{cases} 
      -\operatorname{grad} f_1(\boldsymbol{W}_\ell), & \ell = 0 
      \\ 
      -\operatorname{grad} f_1(\boldsymbol{W}_{\ell}) + \beta_{\ell} \Pi_{\mathcal{T}_{\boldsymbol{W}_{\ell}}^{}}(\boldsymbol{D}_{\ell-1}),\hspace{-1.5mm} & \operatorname{otherwise} 
    \end{cases}
\end{equation}
where $\beta_{\ell}$ is the \gls{cg} parameter at the $\ell$ iteration. We should notice that $\operatorname{grad}f(\boldsymbol{W}_\ell)$ belongs to tangent space $\mathcal{T}_{\boldsymbol{W}_{\ell}}$ 
 and $\boldsymbol{D}_{\ell-1}$ belongs to $\mathcal{T}_{\boldsymbol{W}_{\ell-1}}$. Therefore, they
 cannot be added together in a straightforward way. To this end, we should utilize $\Pi_{\mathcal{T}_{\boldsymbol{W}_{\ell}}}(\cdot)$, which
 denotes the transport mapping operation needed to add/subtract points on different tangent spaces. Here, the Fletcher–Reeves rule is adopted to update $\beta$, which is expressed as \cite{GVK502988711}:
\begin{equation} \label{eq:beta}
    \beta_{\ell} = \frac{\| \operatorname{grad}f_1(\boldsymbol{W}_{\ell}) \|^2_F}{\|\operatorname{grad}f_1(\boldsymbol{W}_{\ell-1}) \|^2_F}.
\end{equation}

After the search direction $\boldsymbol{D}_{\ell}$ is established at $\ell$-th iteration, the beamforming matrix $\boldsymbol{W}$ can be updated as:
\begin{equation} \label{eq:nextpoint}
    \boldsymbol{W}_{\ell+1} = \mathcal{R}\left( \boldsymbol{W}_{\ell} + \alpha_{\ell} \boldsymbol{D}_{\ell} \right),
\end{equation}
where $\mathcal{R}$ is the \textit{retraction} operator which is employed to map the point in the tangent space $\mathcal{T}_{\boldsymbol{W}_\ell}$ back onto the $\mathcal{OB}$, and $\alpha_\ell$ is the step size at the $\ell$-th iteration. $\alpha_k$ is selected by obeying the Wolf conditions \cite{Sato03042015}:
\begin{subequations}
\begin{align}
        &f_1(\boldsymbol{W}_{\ell+1}) \leq f_1(\boldsymbol{W}_\ell) + c_1 \alpha_\ell  \left< f_1(\boldsymbol{W}_\ell),\boldsymbol{D}_\ell\right> \label{eq:armijo}
        \\
        &\left|\left< \operatorname{grad} f_1(\boldsymbol{W}_{\ell+1}), \Pi_{\mathcal{T}_{\boldsymbol{W}_{\ell+1}}}(\boldsymbol{D}_\ell) \right> \right|  \nonumber 
        \\
        & \hspace{35mm} \leq c_2 |\left< \operatorname{grad} f_1(\boldsymbol{W}_{\ell}), \boldsymbol{D}_\ell \right>| , \label{eq:wolf2}
\end{align} 
\end{subequations}
where inequality \eqref{eq:armijo} is the well-known Armijo condition \cite{GVK502988711}. By ensuring the Wolf conditions, the objective function is non-increasing. The retraction operator for the $\mathcal{OB}$ can be computed as
\begin{equation} \label{eq:retraction}
    \mathcal{R}(\boldsymbol{W}) = \rho(\boldsymbol{W}\boldsymbol{W}^H \odot \mathbf{I}_{M_T})^{-1/2} \boldsymbol{W}.
\end{equation}

With these operations, \gls{rcg} proceeds iteratively and is expected to converge to at least a sub-optimal solution of \eqref{eq:SGCDF-RMO1}. The optimization steps of $\boldsymbol{W}$ for solving $\mathcal{SP}_I^{LC}$ are summarized in \textbf{Algorithm \ref{alg:SP1:RMO-SGCDF}}.

\begin{algorithm}[!ht]
\caption{\hspace{-.1cm}{\bf :}  Proposed \gls{rmo}-\gls{sgcdf} for $\mathcal{SP}_{I}^{LC}$}\label{alg:SP1:RMO-SGCDF}
\begin{algorithmic}[1] 
\State {\bf Input:} $R_{\min}$, $\sigma^2$, $\epsilon$, $\boldsymbol{H}_k~\forall k$,
\Statex \textit{Subproblem I: optimal sensing beam pattern}
\State Set $\ell = 0$
\State Initialize solution $\boldsymbol{W}_\ell$ as described in \ref{sss:initial}
\State Obtain $\operatorname{grad}f_1(\boldsymbol{W}_\ell)$ by Eq. \eqref{eq:Rgrad1}
\State Compute search direction $\boldsymbol{D}_\ell$ by Eq. \eqref{eq:descent_direction}
\State Update $\boldsymbol{W}_{\ell+1}$ by Eq. \eqref{eq:nextpoint}

\While{$| f_1(\boldsymbol{W}_{\ell+1}) - f_1(\boldsymbol{W}_\ell) |_F \geq \epsilon$}
\State $\ell = \ell + 1$
\State  Obtain $\operatorname{grad}f_1(\boldsymbol{W}_\ell)$ by Eq. \eqref{eq:Rgrad1}
\State Compute $\beta_\ell$ by Eq. \eqref{eq:beta}
\State Compute search direction $\boldsymbol{D}_\ell$ by Eq. \eqref{eq:descent_direction}
\State Update $\boldsymbol{W}_{\ell+1}$ by Eq. \eqref{eq:nextpoint}
\EndWhile

\State Let us denote $\boldsymbol{\overline{W}}^{\star}$ as the obtained solution for $\mathcal{SP}_I^{LC}$

\State {\bf Output:} DF-Beamforming $\boldsymbol{\overline{W}}^{\star}$ matrix
\end{algorithmic}
\end{algorithm}

\subsubsection{Subproblem II}

In order to solve $\mathcal{SP}_{II}$ \eqref{eq:SP2} in a low-complexity way, we also adopt the \gls{rcg} algorithm similarly to the discussion formerly. It is noteworthy that directly tackling the formulation in \eqref{eq:SP2} would require additional techniques to handle the communication constraint \eqref{eq:c2}, such as the augmented Lagrangian method, which relies on penalty-based mechanisms \cite{10737380,10762897}. Aiming to further reduce the complexity of the proposed methodology and design a feasible communication beam pattern, we extend a projection-based minimization approach over a closed convex set framework, proposed by \cite{10955684}. In contrast to the original formulation in Euclidean space, we extend this idea to the Riemannian manifold setting, enabling the elimination of per-antenna power constraint, represented by the $\mathcal{OB}$, in the set of projections, as done in \cite{10955684}, since \gls{rmo} explores the underlying manifold geometry. Towards this end, we should notice that the communication constraint \eqref{eq:c2} can be written as 
\begin{equation} \label{eq:RateConstraint2}
    \sqrt{\Gamma_k} | \boldsymbol{w}_k^H \boldsymbol{h}_k| \geq \left(\sum_{j=1}^{K+M_T} |\boldsymbol{w}_j \boldsymbol{h}_k|^2 + \sigma^2 \right)^{1/2}, \quad \forall k.
\end{equation}

For the sake of notational simplicity, we introduce the following definitions:
\begin{align}
    \boldsymbol{H}_k &=\hspace{-1mm}  
    \begin{bmatrix}
        \boldsymbol{\tilde{H}}_k \,\mathbf{0}_{M_T(K+M_T)\times 1} \, \boldsymbol{\bar{h}}_k
    \end{bmatrix}^H\hspace{-3mm}, 
    \hspace{-2mm}&  &\in \mathbb{C}^{(K+M_T+2) \times M_T(K+M_T)},
    \\
    \boldsymbol{z} &= 
    \begin{bmatrix}
        \mathbf{0}_{1\times (K+M_T)} ~ \sigma^2 ~ 0
    \end{bmatrix}^T\hspace{-3mm}, 
    &  &\in \mathbb{C}^{(K+M_T+2) \times 1},
\end{align}
where
\begin{equation}
    \boldsymbol{\tilde{H}}_k = \mathbf{I}_{K+M_T} \otimes \boldsymbol{h}_k, \quad \boldsymbol{\bar{h}}_k = \mathbf{e}_k \otimes \sqrt{\Gamma_k}\boldsymbol{h}_k,
\end{equation}
and $\mathbf{e}_k$ denotes the $k$-th column vector of identity matrix $\mathbf{I}_{K+M_T}$. With the above notation, Eq. \eqref{eq:RateConstraint2} is feasible if the following equation is satisfied:
\begin{equation} \label{eq:x}
    \boldsymbol{x}_k(\boldsymbol{W}) = \boldsymbol{H}_k \operatorname{vec}(\boldsymbol{W}) + \boldsymbol{z}, \quad \in \mathcal{C}^k,
\end{equation}
where $\mathcal{C}^k$ is the \gls{soc} expressed as
\begin{equation}
    \mathcal{C}^k = \biggl \{\boldsymbol{x}_k \in \mathbb{C}^{K+M_T+2} \Big| | \tilde{x}_k |\geq  \|\boldsymbol{\bar{x}}_k \|  \biggl\},
\end{equation}
with $\boldsymbol{\bar{x}}_k \in \mathbb{C}^{K+M_T+1}$ being a sub-vector of $\boldsymbol{x}_k$ obtained by removing the last element $\tilde{x}_k \in \mathbb{C}$, i.e., $\boldsymbol{x}_k = [\boldsymbol{\bar{x}}_k^H~ \tilde{x}_k^*]^H$.

Based on this formulation, we can define the following unconstrained problem to obtain a feasible communication beam pattern:
\begin{equation} \label{eq:SGCDF-RMO2}
     \mathcal{SP}_{II}^{LC}: \hspace{-0.5mm} \underset{\boldsymbol{{W}} \in \mathcal{OM}}{\operatorname{min}}  \, f_2(\boldsymbol{W}) \triangleq  \sum_{k=1}^{K} \Big\| \boldsymbol{x}_k(\boldsymbol{{W}}) -  \Pi_{\mathcal{C}^k}\bigl(\boldsymbol{x}_k(\boldsymbol{{W}})\bigl) \Big\|^2,
\end{equation}
which can be interpreted as a projection problem onto a closed convex set, where $\Pi_{\mathcal{C}^k}(\cdot)$ denotes the projection into the $k$-th \gls{soc}, which can be computed in closed-form as
\begin{equation}
    \Pi_{\mathcal{C}^k}(\boldsymbol{x}_k ) \triangleq
    \begin{cases} 
        \boldsymbol{x}_k, & \text{if } \left\| \boldsymbol{\bar{x}}_k \right\|_2 \leq |\tilde{x}_{k}| \\ 
        \mathbf{0}, & \text{if } \left\| \boldsymbol{\bar{x}}_k \right\|_2 \leq -|\tilde{x}_{k}| \\ 
        \boldsymbol{y}, & \text{otherwise}
    \end{cases}
\end{equation}
where $\boldsymbol{y}$ is given as
\begin{equation}
    \boldsymbol{y} = \frac{1}{2}\left(\|[\boldsymbol{\bar{x}}_k] \|+|\tilde{x}_{k}|    \right) \left[
        \frac{\boldsymbol{\bar{x}}^H_k}{\|\boldsymbol{\bar{x}}_k \|} ~
        \frac{\tilde{x}^*_{k}}{|\tilde{x}_{k}|}\right]^H.
\end{equation}

Moreover, the Euclidean gradient of the objective function in problem \eqref{eq:SGCDF-RMO2} can be directly computed utilizing \cite{sun2006short} and \cite[Proposition 1]{10955684}: 
\begin{align}
    \nabla f_2(\boldsymbol{W}) &=  2\operatorname{unvec}\left( \sum_{k=1}^{K}\boldsymbol{H}_k^H \boldsymbol{v}_k(\boldsymbol{W}) \right), \label{eq:Egrad2}
    \\
    \boldsymbol{v}_k(\boldsymbol{W}) &= \boldsymbol{x}_k(\boldsymbol{W}) -\Pi_{\mathcal{C}^k}\Big(\boldsymbol{x}_k(\boldsymbol{W})\Big), \label{eq:v}
\end{align}
where $\operatorname{unvec}(\cdot)$ denotes the inverse vectorization operator, which reshapes a vector of size $M_T(K+M_T)\times1$ into a matrix of size $M_T\times (K+M_T)$. 
The Riemmanian gradient of $f_2(\boldsymbol{W})$ is given as
\begin{align} \label{eq:Rgrad2}
    \operatorname{grad} f_2(\boldsymbol{W}) &= \Pi_{\mathcal{T}_{\boldsymbol{W}}}(\nabla f_2(\boldsymbol{W})) 
    \\
    &= \nabla f_2(\boldsymbol{W}) - \frac{1}{\rho^2}\left(\nabla f_2(\boldsymbol{W}) \boldsymbol{W}^H \odot \mathbf{I}_{M_T}\right) \boldsymbol{W} \nonumber.
\end{align}

%By utilizing the aforementioned \gls{rcg} for solving $\mathcal{SP}_{II}^{LC}$, we can reach a feasible communication waveform. 
The steps for solving $\mathcal{SP}_{II}^{LC}$ by utilizing the aforementioned \gls{rcg} are summarized in \textbf{Algorithm \ref{alg:SP2:RMO-SGCDF}}.

\begin{algorithm}[!ht]
\caption{\hspace{-.1cm}{\bf :} Proposed \gls{rmo}-\gls{sgcdf} for $\mathcal{SP}_{II}^{LC}$}\label{alg:SP2:RMO-SGCDF}
\begin{algorithmic}[1] 
\State {\bf Input:} $\boldsymbol{H}_k~\forall k$, $R_{\min}$, $\sigma^2$, $\epsilon$
\Statex \textit{Subproblem II: communication-feasible beam pattern}

\State Compute $[R_1, R_2,\dots,R_K]$ for $\boldsymbol{\overline{W}}^\star$ as Eq. \eqref{eq:rate}
\If{$\min([R_1, R_2,\dots,R_K]) \geq R_{\min}$}
    \State $\boldsymbol{W}^{\star} = \boldsymbol{\overline{W}}^\star$ 
\Else
    \State Set $\ell = 0$
    \State Initialize solution $\boldsymbol{\overline{W}}_\ell = \boldsymbol{\overline{W}}^\star$
    \State Obtain $\operatorname{grad}f_2(\boldsymbol{\overline{W}}_\ell)$ by Eq. \eqref{eq:Rgrad2}
    \State Compute search direction $\boldsymbol{D}_\ell$ by Eq. \eqref{eq:descent_direction} for $f_2$
    \State Update $\boldsymbol{\overline{W}}_{\ell+1}$ by Eq. \eqref{eq:nextpoint}
    \State Compute $[R_1, R_2,\dots,R_K]$ for $\boldsymbol{\overline{W}_\ell}$ as Eq. \eqref{eq:rate}\While{$\operatorname{min}\left([R_1,R_2,\dots,R_{K}]\right) < R_{\min}$}
    \State $\ell = \ell + 1$
    \State  Obtain $\operatorname{grad}f_2(\boldsymbol{\overline{W}}_\ell)$ by Eq. \eqref{eq:Rgrad2}
    \State Compute $\beta_\ell$ by Eq. \eqref{eq:beta}
    \State Compute search direction $\boldsymbol{D}_\ell$ by Eq. \eqref{eq:descent_direction} for $f_2$
    \State Update $\boldsymbol{\overline{W}}_{\ell+1}$ by Eq. \eqref{eq:nextpoint}
    \State Compute $[R_1, R_2,\dots,R_K]$ for $\boldsymbol{\overline{W}}_{\ell+1}$ as Eq. \eqref{eq:rate}
\EndWhile
\EndIf

\State {\bf Output:} DF-Beamforming $\boldsymbol{W}^{\star}$ matrix
\end{algorithmic}
\end{algorithm}

\subsection{Selection of Initial Point} \label{sss:initial}

Here we propose a suitable initial solution for the \gls{rmo}-\gls{sgcdf} methodology. For sensing, we initialize omnidirectional as $\boldsymbol{W}_S = \sqrt{\frac{p_S}{M_T}}\mathbf{I}_{M_T},$
where $p_S$ denotes the transmit power allocated for sensing. This choice ensures uniform energy distribution across all transmit antennas. Moreover, regarding communication, we adopt the \gls{zf} precoder for initialization, $\boldsymbol{W}_{C}^{\operatorname{ZF}}$, whose the $k$-th column is given as $\boldsymbol{w}^{\operatorname{ZF}}_{c,k} = \sqrt{p_k} \frac{\boldsymbol{v}_k}{\| \boldsymbol{v}_k \|^2}$ where $\boldsymbol{v}_k$ is the $k$-th column of matrix $\boldsymbol{V}$ computed as
$\boldsymbol{V} = (\boldsymbol{H} \boldsymbol{H}^H)^{-1}\boldsymbol{H},$
and $\boldsymbol{p} = [p_1,p_2,\dots,p_K]^T$, is the power allocation vector across users. Here, we set $p_k$ so that all users satisfy the minimum rate of communication, $R_{\min}$. To achieve this condition, we have adapted the technique proposed in \cite{1406483,7031971} for our scenario, and this equal-rate condition is met if $\boldsymbol{p}$ is computed as \footnote{Notice that the feasibility of \eqref{eq:power}, i.e., $p_k \geq 0,~\forall k$, is achieved only if $|\lambda_{\max}(\boldsymbol{\Delta})|\leq 1$ \cite{1406483}, which obviously depends on values of $\boldsymbol{H}$, and $R_{\min}$.}
\begin{equation} \label{eq:power}
    \boldsymbol{p} =  \boldsymbol{\Delta}^{-1} \Bigg(\sigma^2\mathbf{I}_K +\left( \boldsymbol{H}^H\boldsymbol{W}_s \boldsymbol{W}_s^H \boldsymbol{H} \right)\odot \mathbf{I}_K\Bigg)\mathbf{1}
_{K},
\end{equation}
where $\mathbf{1}_K$ is the $K$-size vector of ones, while $\boldsymbol{\Delta} \in \mathbb{R}^{K\times K}$ is a matrix whose elements are computed as
\begin{equation}
    \left[\boldsymbol{\Delta}\right]_{j,i} = 
    \begin{cases} 
        \frac{\left| \boldsymbol{v}_i^H \boldsymbol{h}_i \right|^2}{(2^{R_{\min}} - 1) \| \boldsymbol{v}_i \|^2} & \text{for } i = j, \\ - \frac{|\boldsymbol{v}_i^H \boldsymbol{h}_j|^2}{\|\boldsymbol{v}_i\|^2} 
         & \text{for } i \neq j.
    \end{cases}
\end{equation}

To ensure that the initial point belongs to $\mathcal{OM}$, the retraction operator \eqref{eq:retraction} is applied.

\begin{remark}
    By initializing $p_S = 0$, the sensing precoder is effectively deactivated, thereby the \gls{rmo}-\gls{sgcdf} yields that $\boldsymbol{W}^\star = \boldsymbol{W}_C^\star$, which obviously reduces \gls{dof}. In this case, the system operates without a dedicated sensing stream. This corresponds to a particular scenario of our proposed framework, where the sensing task is entirely carried out using only communication signals, and has been adopted in different works 
    \cite{10737380,9591331,10888351,9667503,9805471}. This demonstrates the robustness of our proposed \gls{rmo}-\gls{sgcdf} methodology by covering different practical scenarios. 
\end{remark}

\section{Convergence and Computational Complexity Analysis of RMO-SGCDF}

In this section, we provide the convergence analysis and computational complexity of the proposed algorithms in terms of \glspl{flop}.

\subsection{Convergence Analysis of RMO-SGCDF Algorithm}

We aim to establish the convergence of the proposed \gls{rmo}-\gls{sgcdf} algorithm to a local minimizer. Since the algorithm is based on the \gls{rcg} method, it suffices to demonstrate that the Riemannian gradient converges to zero. To this end, we present the following theorem.

\begin{theorem} \label{theo:zouten}
    Zoutendijk Condition: Consider any iteration of form \eqref{eq:nextpoint}, where $\alpha_k$ satisfies the strong Wolf conditions \eqref{eq:armijo}, \eqref{eq:wolf2}, where $\operatorname{grad} f_i(\cdot)$ is limited below, i.e., $f_i(\boldsymbol{W}_\ell) \leq f_i(\boldsymbol{W}_0)$ $\forall \ell,i$, where $\boldsymbol{W}_0$ is the starting point of the iteration. Let $\phi_\ell$ the angle between $\operatorname{grad} f_i(\boldsymbol{X}_\ell)$ and the descent direction $\boldsymbol{D}_\ell$ as 
    \begin{equation} \label{eq:cos}
        \cos(\phi_{i,\ell}) = -\frac{\left< \operatorname{grad} f_i(\boldsymbol{W}_\ell) , \boldsymbol{D}_\ell \right> }{\| \operatorname{grad} f_i(\boldsymbol{W}_\ell) \|_F \| \boldsymbol{D}_\ell \|_F}.
    \end{equation}
    Assume also that the gradient $\operatorname{grad} f_i$, retraction operator $\mathcal{R}(\cdot)$, and projection onto tangent plane $\Pi_{\mathcal{T},\boldsymbol{W}}(\cdot)$ are Lipschitz continuous. Then 
    \begin{equation} \label{eq:zouten}
        \sum_{\ell=0}^{\infty} \cos^2(\phi_{i,\ell}) \| \operatorname{grad} f_i( \boldsymbol{W}_\ell) \|^2 \leq \infty.
    \end{equation}
    \begin{proof}
        The proof is omitted for brevity.
    \end{proof}
\end{theorem}

Next, we utilize the following Lemma, which shows that the inner product of the Riemannian gradient $\operatorname{grad} f_i(\boldsymbol{W}_\ell)$ and the conjugate direction $\boldsymbol{D}_\ell$ is bounded.
\begin{lemma}
    If the steep length $\alpha_\ell$ satisfies the strong Wolf conditions \eqref{eq:armijo}, \eqref{eq:wolf2}, with $c_2 < 1/2$, then for every $\ell \in \mathbb{N}$
    \begin{equation}\label{eq:lemma1}
        -\frac{1}{1-c_2} \leq \frac{\left< \operatorname{grad} f_i(\boldsymbol{W}_\ell), \boldsymbol{D}_\ell \right>}{\| \operatorname{grad} f_i(\boldsymbol{W}_\ell) \|^2_F} \leq \frac{2c_2-1}{1-c_2}
    \end{equation}
    \begin{proof}
        Please refer to \cite[Lemma 4.1]{Sato03042015}.
    \end{proof}
\end{lemma}
Finally, Theorem \ref{theo:convergence} shows that the \gls{rcg} iteration of the proposed scheme converges to a local minimizer $\boldsymbol{W}^*$ of $f_i(\boldsymbol{W}),~\forall i =1,2$ on the manifold $\mathcal{OM}(M_T,K+M_T)$.
\begin{theorem} \label{theo:convergence}
    Let $\boldsymbol{W}_\ell \in \mathcal{OM}$ a point and $\boldsymbol{D}_\ell \in \mathcal{T}_{\boldsymbol{W}_\ell}$ the descent direction at the $\ell$ iteration, with the Fleetcher-Reeves rule $\beta_\ell = \frac{\| \operatorname{grad} f_i(\boldsymbol{W}_\ell) \|^2}{\|\operatorname{grad} f_i (\boldsymbol{W}_{\ell-1}) \|^2}$. If the steep length $\alpha_\ell$ satisfies the strong Wolf conditions \eqref{eq:armijo}, \eqref{eq:wolf2}, with $c_2 < 1/2$ then
    \begin{equation} \label{eq:convergence}
        \lim_{\ell\rightarrow \infty} \operatorname{inf} \| \operatorname{grad} f_{i}(\boldsymbol{W}_\ell) \|^2_{\operatorname{F}} = 0, \quad i=\{1,2\},
    \end{equation}
    \begin{proof}
        The proof is omitted due to the sake of space.
    \end{proof}
\end{theorem}

\subsection{Analysis of Computational Complexity}

\subsubsection{SGCDF}
We first provide the complexity analysis of \textbf{Algorithm \ref{alg:SGC}}. The computational cost of the proposed \gls{sgcdf} depends on the two subproblems, namely $\mathcal{SP}{I}$ and $\mathcal{SP}{II}$. For a given solution accuracy $\epsilon$, the worst-case complexity of solving $\mathcal{SP}_{I}$ using the primal–dual interior-point method is of $\mathcal{O}((M_{T}+T)^{6.5} \log(1/\epsilon))$ \cite{10050406}, while $\mathcal{SP}_{II}$ is a quadratic \gls{sdp}, and its complexity is of $\mathcal{O}(K^{6.5}M_{T}^{6.5} \log(1/\epsilon))$ \cite{Liu2020}. Therefore, the total complexity is of $\mathcal{O}( ((M_T +T)^{6.5} + K^{6.5}M_{T}^{6.5} )\log(1/\epsilon))$.
\subsubsection{RMO-SGCDF}
We now analyze the computational complexity of the \gls{rmo}-\gls{sgcdf} algorithm, beginning with $\mathcal{SP}_I^{LC}$. In the first step of \textbf{Algorithm~\ref{alg:SP1:RMO-SGCDF}}, the Riemannian gradient in \eqref{eq:Rgrad1} is computed with complexity of $\mathcal{O}(T^4 +M_T^2(T^3 + M_T + K))$. Using this gradient, the computation of the \gls{cg} parameter $\beta$ in \eqref{eq:beta} has complexity of $\mathcal{O}(M_T(K+M_T))$. Next, updating the Riemannian conjugate direction $\boldsymbol{D}_\ell$ in \eqref{eq:descent_direction} has complexity of $\mathcal{O}(M_T^2(K+M_T))$. The step size $\alpha_\ell$ is obtained via line search, where the dominant cost comes from the Riemannian gradient evaluation, i.e., $\mathcal{O}(M_T^3 + M_T^2(K+T^3))$. Finally, updating $\boldsymbol{W}$ as in \eqref{eq:nextpoint} has complexity of $\mathcal{O}(M_T^2(K+M_T))$. Therefore, the total complexity is of $\mathcal{O}(M_T^3 + M_T^2(K+T^3))$. Regarding $\mathcal{SP}_{II}^{LC}$, the main difference lies in the objective function, whose evaluation $KM_T(K+M_T)^2$ computations. For clarity, Table~\ref{tab:complexity} summarizes the computational complexity of the main algorithms.
\renewcommand{\arraystretch}{1.3}
{\tiny
\begin{table}[!h]
\centering
\caption{Comparison of Computational Complexity}
\label{tab:complexity}
\begin{tabular}{|p{1.62cm}|p{6cm}|}
\hline
\textbf{Algorithm} & \textbf{Complexity} \\
\hline
\textit{\gls{sgcdf}} & 
$\mathcal{O}\big(((M_T + T)^{6.5} + K^{6.5} M_T^{6.5}) \log(1/\epsilon)\big)$ \\
\hline
\textit{\gls{rmo}-\gls{sgcdf}} & 
$\mathcal{O}\big(I_{1} (T^4 +M_T^{2}(M_T + K + T^3)) + I_2(K^3M_T + K^2M_T^2 + KM_T^3\big)$ \\
\hline
\textit{Comm\&Sensing} & 
$\mathcal{O}\big((K M_T + T)^{6.5} \log(1/\epsilon)\big)$ \\
\hline
\textit{Sensing only} & 
$\mathcal{O}\big((M_T + T)^{6.5} \log(1/\epsilon)\big)$ \\
\hline
\end{tabular}
\end{table}
}

\renewcommand{\arraystretch}{1.0}
\section{Simulation Results and Discussion}\label{sec:results}

We provide numerical results to evaluate the performance of the proposed algorithms \textbf{“\gls{sgcdf}”} and \textbf{“\gls{rmo}-\gls{sgcdf}”} for beam-pattern design.

\subsection{Simulation Setup and Parameters}

We assume $T=3$ targets, with deterministic fixed positions where the angles of the three targets are set as $\theta_1=-45^\circ$, $\theta_2=30^\circ$, and $\theta_3=60^\circ$. Moreover, we assume different target positions, $\rho_1 = 50~m$, $\rho_2 = 60~m$, and $\rho_3 = 70~m$. The path-loss model is given by $\beta(d) = C_0(\frac{d_0}{d_i})^{\lambda_i}$, where $C_0 = - 30$ dB represents the path-loss at the reference distance $d_0=1m$, $d_i$ represents the individual link distance, and $\lambda_i$ denotes the path-loss exponent.  Unless otherwise stated, the remaining simulation parameters are as in Table \ref{table:1}.  

\begin{table}[htbp] 
\centering
\caption{Simulation and Algorithmic Parameters}
\begin{tabular}{|l|l|}
\hline
\textbf{Parameters} & \textbf{Values} \\
\hline
Noise variance: & $\sigma^2 =-96$ dBm 
\\
Distance between elements: & $d = \lambda/2$ 
\\
Carrier frequency: & $f=6$ GHz   
\\
Number of communication users: & $K=6$ 
\\
Number of snapshots: & $L=1024$
\\
Number of targets: & $T=3$ 
\\
Number of transmit antennas: &  $M_T = 32$   \\
Number of received antennas:
& $M_R = 32$  \\
Rician factor: & $\kappa = 0.1$
\\

Algorithmic convergence parameter: & $\epsilon = 10^{-3}$  \\
\hline
\end{tabular}
\label{table:1}
\end{table}

To establish a centralized communication point, the \gls{bs} is positioned at coordinates $(0,0)$, and we consider a simulation area defined by a semicircular region with radius in the range $\rho_{\min} = 50 ~m$, $\rho_{\max} = 55 ~m$, and angular span from $\theta_{\min} = -25^\circ$ to 
$\theta_{\max} = 25^\circ$,  within which $K=6$ \glspl{ue} are randomly placed. The results are averaged over 100 generated independent random realizations.

\subsection{Benchmarks Schemes}

For comparison purposes, we adopt the following five baseline schemes:
\begin{enumerate}
    \item \textit{Omnidirectional:} This benchmark intends to emit equal power along all different angles, by setting the full orthogonal beamforming matrix. 
    %as $\boldsymbol{W}_C = \begin{bmatrix}
%\mathbf{I}_{K}\sqrt{p_C/K}  ~
% \mathbf{0}_{M_T-K,K} \\
%\end{bmatrix}^T$ and $\boldsymbol{W}_S = \mathbf{I}_{M_T}\sqrt{P_S/M_T}$. We attribute equal power allocation for both functionalities, i.e., $p_S = p_C = P_{\max}/2$.
    \item \textit{Sensing only:} This benchmark utilizes the standard \gls{sdr} optimization technique for solving problem \eqref{eq:problemSDPrelax} without accounting for the communication constraint.

    \item \textit{Comm\&Sensing:} This benchmark utilizes standard the \gls{sdr} optimization technique for solving problem \eqref{eq:problemSDPrelax} accounting the communication constraint.

    \item \textit{ZF\&Sensing:} This benchmark uses the conventional \gls{zf} as fixed \gls{ue} precoding while only optimizing the precoding for sensing and the \gls{ue} power allocation by solving a convex problem, as detailed in \cite[Section III-B]{10756646}.

    \item \textit{C\&S Robust:} This benchmark utilizes the standard \gls{sdr} optimization technique for solving the joint communication and sensing with min-max eigenvalue formulation \cite{10380213}.
\end{enumerate}

All convex problems are solved using the CVX tool with MATLAB.

\subsection{Minimum Communication Rate Setting}

To enable a fair comparison among all considered benchmarks, the minimum communication rate is defined the same for all communication users as $R_{\min}$ and must be appropriately set. To  ensure the feasibility of optimization problems across all independent random realizations of the channel matrix $\boldsymbol{H}$, in this work, we set $R_{\min}$ to be proportional to the achievable rate obtained by solving the max-min fairness problem under the \gls{zf}, which is known to be near-optimal in the high \gls{snr} regime.
\begin{equation}
    R_{\min} = \delta R_{\max}^{\operatorname{ZF}},
\end{equation}
where $\delta \in [0,1]$ is a scalar referred to as the overload factor.

\subsection{Beam Pattern Analysis}
 
\begin{figure}[ht!]
    \centering
    \includegraphics[width=1\linewidth]{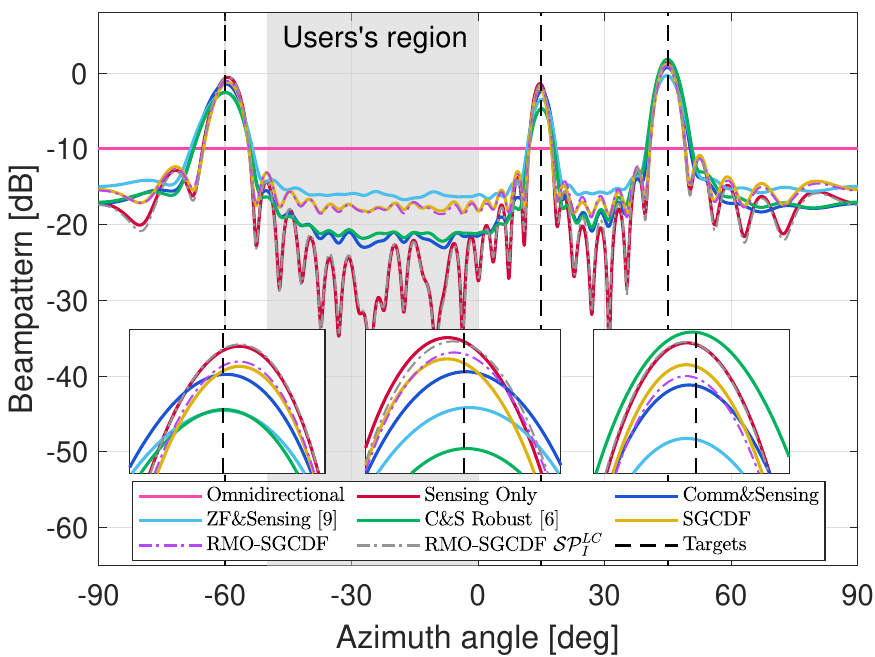}
    \caption{Beampattern for different schemes with $P_{\max} = 20$ dBm, $\delta = 0.7$, for all $\theta \in [-90^\circ,90^\circ]$. } 
    \label{fig:beampattern}
\end{figure}

Fig. \ref{fig:beampattern} illustrates the obtained optimized beam pattern for different strategies, with $P_{\max} = 20$ dBm.  As observed, all evaluated beam patterns effectively steer their main lobes toward the target directions. Besides, we can see from this figure that, as expected, the “Sensing only” scheme produces the most adequate beam pattern for sensing purposes, exhibiting the highest power gains in the directions of the targets. This gain is possible because this scheme does not account for the constraint of communication, thereby enabling more power to be irradiated toward the targets, while minimizing the power radiated in the direction of the communication users. 

Proceeding, we can see that “Comm\&Sensing”  aligns the beam pattern toward the targets, while satisfying the communication constraint, i.e., resulting in a dual-function beam pattern. As a result, there is a noticeable reduction in power emitted toward the targets compared with the former scheme. At the same time, this scheme presented the lowest power irradiated toward the communication user's region compared to the dual-function approaches.  Furthermore, from Fig. \ref{fig:beampattern} we can see that both of the proposed methods strike a favorable trade-off between the power irradiated toward the targets and the communication users.  
Remarkably, both the designs outperform all dual-function benchmarks in terms of beam pattern gain.

The “ZF\&Sensing” and “C\&S Robust” methods show a strong limitation in terms of beam pattern gain. The gain toward all the targets' direction is noticeably limited, since these methodologies imply higher power towards the communication users and the farther target, respectively. These properties are attributed to the reduced \gls{dof} caused by fixing the phase of the communication precoder to the \gls{zf} solution, and the min-max formulation, respectively.

\begin{figure}[ht!]
    \centering
    \includegraphics[width=1\linewidth]{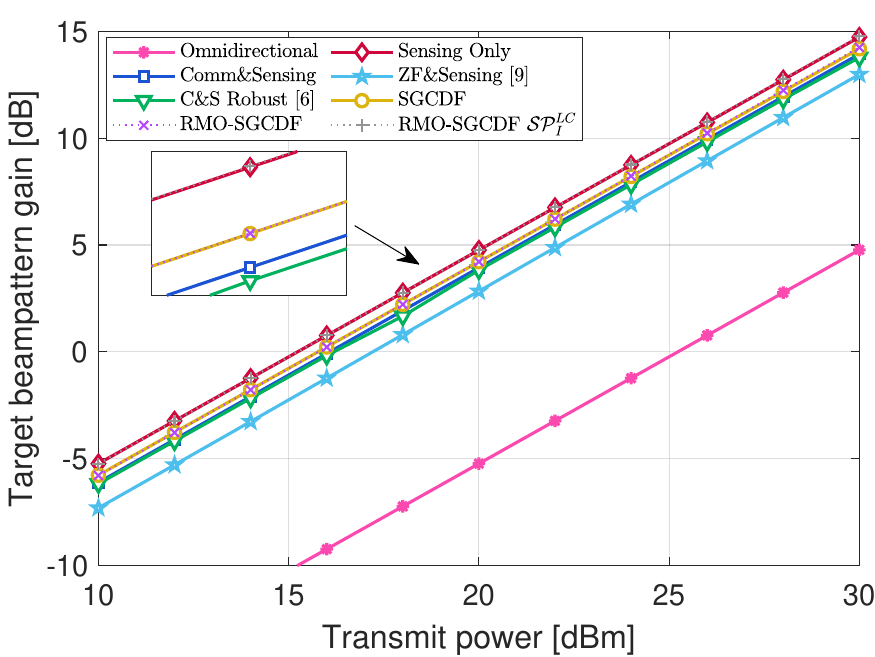}
    \caption{Sum-beampattern gain ($10\log_{10}(\sum_{t=1}^{T} B(\theta_t))$) for different schemes with $\delta = 0.7$.} 
    \label{fig:beampatter_gain}
\end{figure}

Fig. \ref{fig:beampatter_gain} shows the sum of beampattern gain toward all targets, %given by $\sum_{t=1}^{T} B(\theta_t)$, 
as a function of the total transmit power. This figure extends the qualitative observation made in Fig. \ref{fig:beampattern}, confirming that the behavior observed in a fixed-power scenario holds consistently across the range of considered power. This result reveals that regardless of the power value, our proposed methods can achieve higher beam pattern gains, 
achieving performance that closely approaches that of the ideal “Sensing only”. This outcomes is expected, since the proposed \gls{sgcdf} framework  is formulated to explicitly track the physical pattern of the appropriate sensing beam pattern obtained by solving the $\mathcal{SP}_I~(\mathcal{SP}_I^{LC})$, and then projected onto the feasible communication beam pattern set through $\mathcal{SP}_{II}~(\mathcal{SP}_{II}^{LC})$. This two-stage strategy enables the proposed solution to retain most of the desirable sensing characteristics, even in the presence of the communication requirement. 
Moreover, the figure highlights the performance limitations of the “Omnidirectional” and “ZF\&Sensing” schemes, which exhibit significantly lower beam pattern gains across all power levels.

%In the following, we analyze how these differences in beampattern gain impact the accuracy of \gls{doa} estimation, a key sensing functionality, thus further validating the practical benefits of each method.

\subsection{RMSE and RCRLB}

\begin{figure}[ht!]
    \centering
    \includegraphics[width=1\linewidth]{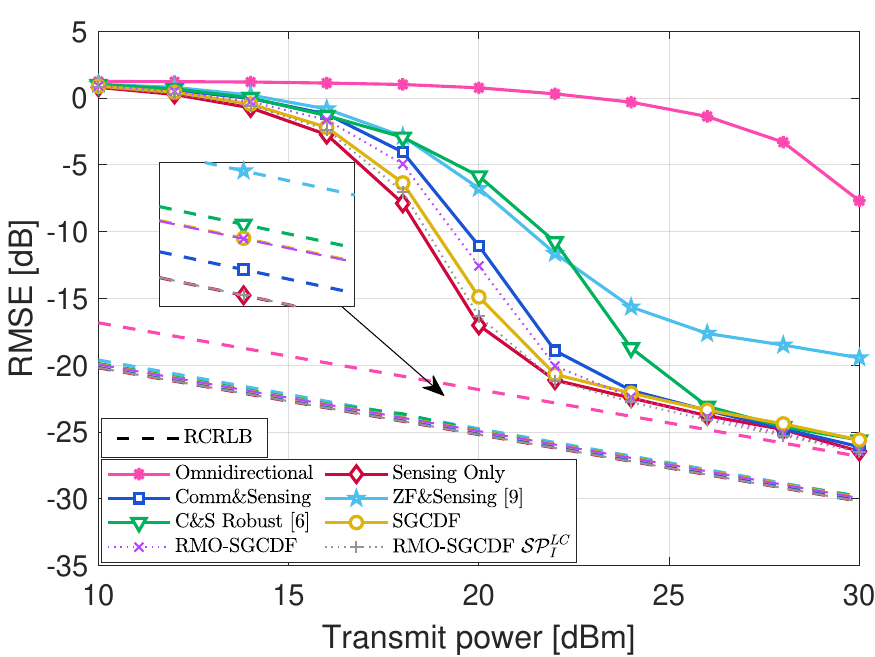}
    \caption{ \gls{rmse} and \gls{rcrlb} for different schemes with $\delta = 0.7$.} 
    \label{fig:RMSE_power}
\end{figure}

Fig. \ref{fig:RMSE_power} presents a comparison of the \gls{doa} estimation performance for the considered schemes, evaluated in terms of \gls{rmse}. Due to the prohibitive complexity of \gls{mle} \cite{10050406,10138058} in multi-target scenarios, in this work, we employ the classical \gls{music} algorithm \cite{1143830,10135096,10251151} to estimate the \glspl{doa} of the targets. 

Firstly, it can be seen from Fig. \ref{fig:RMSE_power} that the gap between the empirical \gls{rmse} and theoretical \gls{rcrlb} decreases as the transmit signal power budget $P_{\max}$ increases. This behavior is expected, as higher transmit power leads to improved \gls{snr}. Moreover, at high transmit power levels, the \gls{rmse} closely approaches the \gls{rcrlb}, which validates the effectiveness of utilizing \gls{crlb} as the performance metric in designing the beam pattern. 

Among the evaluated strategies, the “Sensing only” scheme achieves the lowest \gls{rmse} across all power levels, confirming its superior sensing capability. This result aligns with the beam pattern analysis in Fig. \ref{fig:beampattern}. Additionally, we can see that both proposed methodologies outperform the baseline dual-function designs “Comm\&Sensing”, “ZF\&Sensing”, and “C\&S Robust” in terms of \gls{rmse}.

\begin{comment}
\begin{figure}[ht!]
    \centering
    \includegraphics[width=1\linewidth]{figure6.pdf}
    \caption{\gls{rcrlb} for different schemes with $\delta = 0.7$.} 
    \label{fig:RCRLBbar}
\end{figure}
\end{comment}

\begin{figure}[ht!]
    \centering
    \includegraphics[width=1\linewidth]{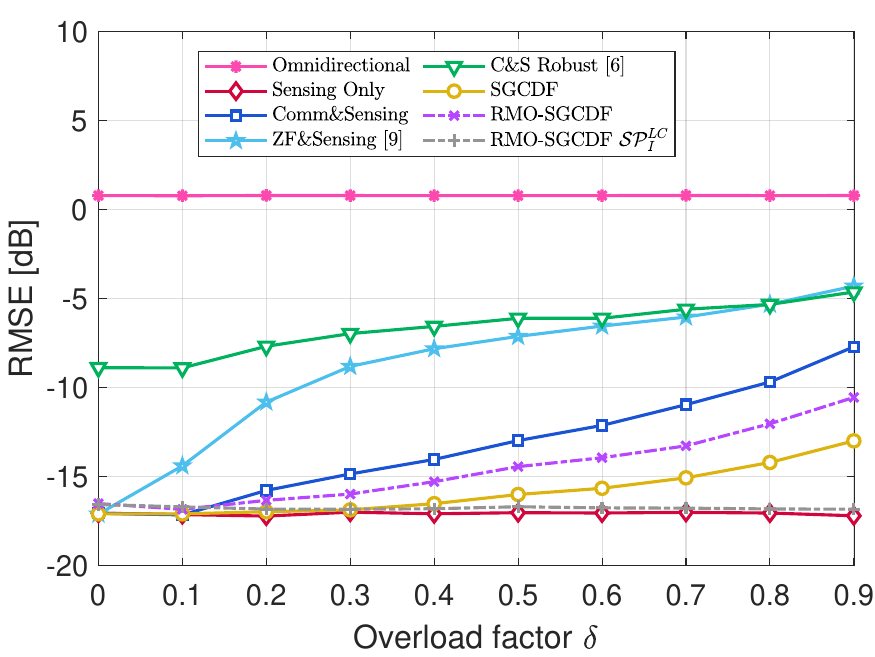}
    \caption{RMSE of different schemes for different overload factors $\delta$ with $P_{\max} = 20$ dBm.} 
    \label{fig:RMSE_overload}
\end{figure}

Interestingly, the “Comm\&Sensing” scheme, despite achieving the lowest \gls{rcrlb} among the dual-function methods, does not yield inferior \gls{rmse} results compared to the proposed strategies. This obsevation reveals an important insight: minimizing the \gls{crlb} by considering the communication constraint does not necessarily guarantee minimized \gls{rmse} in practice. In fact, the formulation of “Comm\&Sensing” directly aims to minimize the \gls{crlb} value while guaranteeing the communication feasibility, which may result in beam patterns that are suboptimal from the \gls{rmse} point of view. 

%\begin{comment}
\begin{figure*}[ht!]
    \centering
    \includegraphics[width=0.92\linewidth]{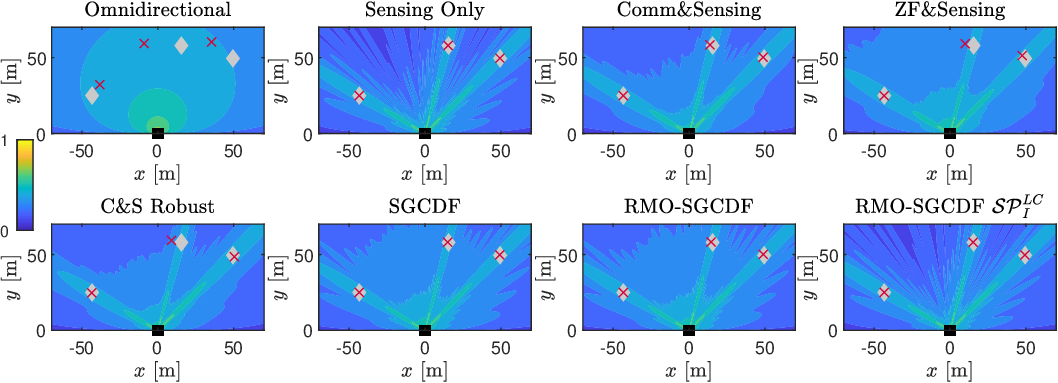}
    \caption{ Normalized average \gls{snr} in dB for the $X=Y$ plane with $P_{\max}=20$ dBm and $\delta = 0.7$. Here, we plot different covariance matrix optimization schemes. For visualization purposes, the normalized \gls{snr} values are truncated at 0 dB. The targets are represented by gray diamonds while their corresponding estimates are denoted in red color, and the \gls{bs} is represented as a black square. } 
    \label{fig:countour}
\end{figure*}
%\end{comment}
This outcome highlights the central motivation behind our design: by decomposing the joint \gls{crlb} minimization problem into two subproblems, we achieve a more effective trade-off between the obtained beam pattern and practical \gls{rmse} estimation, enabling more robust beam patterns. %in achieving accurate \gls{doa} estimation. 

%Finally, we observe that the performance of the “ZF\&Sensing” scheme is significantly affected by fixing the phase of the communication precoding. This limits the flexibility of the waveform design as viewed in Fig. \ref{fig:beampattern}, which, as a result, degrades the sensing performance. This degradation confirms the strong relationship between the amount of power radiated toward the targets, i.e., the obtained waveform, and the resulting \gls{rmse} performance. 

%\begin{comment}
In Fig. \ref{fig:RMSE_overload}, we assess the impact of the system overload on the sensing performance. We can see that by increasing the overload factor,  the sensing performance is  impacted negatively. This performance drop is expected, since higher communication rates require a larger portion of resources. Consequently, fewer resources remain available for shaping the beam pattern to concentrate energy in the target directions. Besides, we can see that our proposed methodology demonstrates strong robustness against the effects of overload.

\subsection{Average Running Time}

\begin{figure}[H]
    \centering
    \includegraphics[width=1\linewidth]{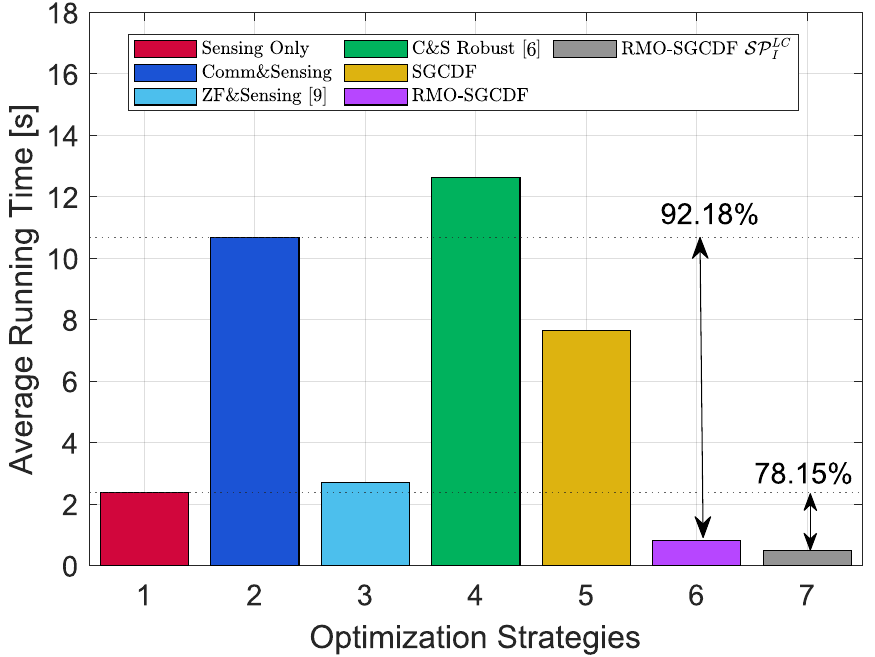}
    \caption{Average elapsed time for each scheme in seconds.} 
    \label{fig:runningTime}
\end{figure}

Fig. \ref{fig:runningTime} compares the average running times of all considered schemes. 
We can see that the “Comm\&Sensing”, “C\&S Robust”, and “SGDFG” algorithms have significantly higher execution times compared to all benchmarks. It can be justified since these methodologies operate in the covariance matrix domain $\boldsymbol{R}_X$, $\{\boldsymbol{R}_{C,k}\}~\forall k$, jointly with the communication constraints. On the other hand,  our proposed \gls{rmo}-\gls{sgcdf} offer a significant complexity reduction, specifically $78.15$, and $92.18\%$ for $\mathcal{SP}_{I}$, and $\mathcal{SP}_{I} +\mathcal{SP}_{II}$, respectively. This result confirms the great trade-off of our proposed method in terms of performance-complexity, as indicated in Table \ref{tab:complexity} in terms of \glspl{flop}, confirming the potential of \gls{rmo}-based techniques.

\subsection{Estimation Performance}

Fig.~\ref{fig:countour} illustrates the impact of each adopted benchmark in a practical target-position estimation scenario. In this evaluation, the \gls{snr} values are computed utilizing the average covariance matrix obtained through 100 realizations, while \gls{doa} for each scheme is obtained by averaging the \gls{doa} obtained by the \gls{music} algorithm. Besides, the distances of the targets are assumed to be perfectly known. The results show that both the proposed methodologies achieve highly accurate target position estimates, performing very close to the dual-function “Comm\&Sensing” method. Besides, the “Sensing only” scheme yields the most accurate results overall, as its beam pattern is optimized exclusively for sensing, with no consideration for communication requirements. 

 \section{Conclusion}\label{sec:conclusion}

A novel framework has been proposed for beam pattern design in a multi-target \gls{mu} \gls{dfmimo} \gls{isac} system, where two \glspl{sp} are properly designed. The first one is formulated to design an appropriate radar beam pattern, and the second guarantees the feasibility of the communication applications. Both of the problems are solved in the covariance domain by traditional convex tools. To further reduce the complexity of the problem, we have considered the problem in the beamforming domain. By modeling the per-antenna constraint as a manifold and the \gls{sinr} constraints as \gls{soc}, we have proposed an efficient solution by utilizing \gls{rmo}. The numerical results illustrate that the proposed schemes outperform the traditional schemes in terms of sensing performance. Besides, the proposed low-complexity algorithm offers  performance similar to the original design at a much lower complexity, and therefore, it offers a favorable performance-complexity trade-off.


\begin{thebibliography}{10}
\providecommand{\url}[1]{#1}
\csname url@samestyle\endcsname
\providecommand{\newblock}{\relax}
\providecommand{\bibinfo}[2]{#2}
\providecommand{\BIBentrySTDinterwordspacing}{\spaceskip=0pt\relax}
\providecommand{\BIBentryALTinterwordstretchfactor}{4}
\providecommand{\BIBentryALTinterwordspacing}{\spaceskip=\fontdimen2\font plus
\BIBentryALTinterwordstretchfactor\fontdimen3\font minus \fontdimen4\font\relax}
\providecommand{\BIBforeignlanguage}[2]{{%
\expandafter\ifx\csname l@#1\endcsname\relax
\typeout{** WARNING: IEEEtran.bst: No hyphenation pattern has been}%
\typeout{** loaded for the language `#1'. Using the pattern for}%
\typeout{** the default language instead.}%
\else
\language=\csname l@#1\endcsname
\fi
#2}}
\providecommand{\BIBdecl}{\relax}
\BIBdecl

\bibitem{10188491}
F.~Liu, L.~Zheng, Y.~Cui, C.~Masouros, A.~P. Petropulu, H.~Griffiths, and Y.~C. Eldar, ``{Seventy Years of Radar and Communications: The road from separation to integration},'' \emph{IEEE Signal Processing Magazine}, vol.~40, no.~5, pp. 106--121, 2023.

\bibitem{Liu2020}
X.~Liu, T.~Huang, N.~Shlezinger, Y.~Liu, J.~Zhou, and Y.~C. Eldar, ``Joint transmit beamforming for multiuser mimo communications and mimo radar,'' \emph{IEEE Transactions on Signal Processing}, vol.~68, pp. 3929--3944, 2020.

\bibitem{11111722}
A.~Magbool, V.~Kumar, Q.~Wu, M.~D. Renzo, and M.~F. Flanagan, ``{A Survey on Integrated Sensing and Communication With Intelligent Metasurfaces: Trends, Challenges, and Opportunities},'' \emph{IEEE Open Journal of the Communications Society}, pp. 1--1, 2025.

\bibitem{9652071}
F.~Liu, Y.-F. Liu, A.~Li, C.~Masouros, and Y.~C. Eldar, ``{Cramér-Rao Bound Optimization for Joint Radar-Communication Beamforming},'' \emph{IEEE Transactions on Signal Processing}, vol.~70, pp. 240--253, 2022.

\bibitem{10955684}
J.~Wang, B.~Wang, J.~Fang, and H.~Li, ``{Low-Complexity Joint Communication and Sensing Beamforming for ISAC Systems: A Bisection Search Approach},'' \emph{IEEE Internet of Things Journal}, vol.~12, no.~13, pp. 25\,620--25\,632, 2025.

\bibitem{10380213}
Z.~Zhao, L.~Zhang, R.~Jiang, X.-P. Zhang, X.~Tang, and Y.~Dong, ``{Joint Beamforming Scheme for ISAC Systems via Robust Cramér–Rao Bound Optimization},'' \emph{IEEE Wireless Communications Letters}, vol.~13, no.~3, pp. 889--893, 2024.

\bibitem{10938377}
Y.~Geng, T.~H. Cheng, K.~Zhong, and K.~C. Teh, ``{Cramér-Rao Bound Minimization for IRS-Aided Multi-User Multi-Target MIMO ISAC Systems},'' \emph{IEEE Transactions on Vehicular Technology}, vol.~74, no.~8, pp. 12\,561--12\,575, 2025.

\bibitem{10097000}
M.~Zhu, L.~Li, S.~Xia, and T.-H. Chang, ``{Information and Sensing Beamforming Optimization for Multi-User Multi-Target MIMO ISAC Systems},'' in \emph{ICASSP 2023 - 2023 IEEE International Conference on Acoustics, Speech and Signal Processing (ICASSP)}, 2023, pp. 1--5.

\bibitem{10756646}
O.~Alp~Topal, O.~Tugfe~Demir, E.~Bjornson, and C.~Cavdar, ``{Multi-Target Integrated Sensing and Communications in Massive MIMO Systems},'' \emph{IEEE Wireless Communications Letters}, vol.~14, no.~2, pp. 345--349, 2025.

\bibitem{11126093}
Z.~Ma, J.~Li, Z.~Liu, G.~Han, T.~Li, and Q.~Guo, ``{Joint Design of Beamforming and IRS Manipulation for Secure Transmission in ISAC System With CRB Constraint},'' \emph{IEEE Wireless Communications Letters}, pp. 1--1, 2025.

\bibitem{10762897}
Y.~Geng, T.~Hiang~Cheng, K.~Zhong, K.~Chan~Teh, and Q.~Wu, ``{Joint Beamforming for CRB-Constrained IRS-Aided ISAC System via Product Manifold Methods},'' \emph{IEEE Transactions on Wireless Communications}, vol.~24, no.~1, pp. 691--705, 2025.

\bibitem{10737380}
S.~Zargari, D.~Galappaththige, C.~Tellambura, and H.~V. Poor, ``{A Riemannian Manifold Approach to Constrained Resource Allocation in ISAC},'' \emph{IEEE Transactions on Communications}, vol.~73, no.~5, pp. 3655--3670, 2025.

\bibitem{9591331}
X.~Wang, Z.~Fei, J.~Huang, and H.~Yu, ``{Joint Waveform and Discrete Phase Shift Design for RIS-Assisted Integrated Sensing and Communication System Under Cramer-Rao Bound Constraint},'' \emph{IEEE Transactions on Vehicular Technology}, vol.~71, no.~1, pp. 1004--1009, 2022.

\bibitem{10411853}
X.~Yang, Z.~Wei, Y.~Liu, H.~Wu, and Z.~Feng, ``{RIS-Assisted Cooperative Multicell ISAC Systems: A Multi-User and Multi-Target Case},'' \emph{IEEE Transactions on Wireless Communications}, vol.~23, no.~8, pp. 8683--8699, 2024.

\bibitem{10888351}
T.~Fang, N.~T. Nguyen, and M.~Juntti, ``{Low-Complexity Cramér-Rao Lower Bound and Sum Rate Optimization in ISAC Systems},'' in \emph{ICASSP 2025 - 2025 IEEE International Conference on Acoustics, Speech and Signal Processing (ICASSP)}, 2025, pp. 1--5.

\bibitem{10065868}
Y.~Mai and H.~Du, ``{Joint Beamforming and Phase Shift Design for RIS-Aided ISAC System},'' in \emph{2022 IEEE 8th International Conference on Computer and Communications (ICCC)}, 2022, pp. 155--160.

\bibitem{9500663}
Z.~Abu-Shaban, K.~Keykhosravi, M.~F. Keskin, G.~C. Alexandropoulos, G.~Seco-Granados, and H.~Wymeersch, ``{Near-field Localization with a Reconfigurable Intelligent Surface Acting as Lens},'' in \emph{ICC 2021 - IEEE International Conference on Communications}, 2021, pp. 1--6.

\bibitem{Song2023}
X.~Song, J.~Xu, F.~Liu, T.~X. Han, and Y.~C. Eldar, ``{Intelligent Reflecting Surface Enabled Sensing: Cramér-Rao Bound Optimization},'' \emph{IEEE Transactions on Signal Processing}, vol.~71, no.~1, pp. 2011--2026, 2023.

\bibitem{Kay97}
S.~M. Kay, \emph{{Fundamentals of Statistical Signal Processing: Estimation Theory}}.\hskip 1em plus 0.5em minus 0.4em\relax Prentice Hall, 1997.

\bibitem{1703855}
I.~Bekkerman and J.~Tabrikian, ``{Target Detection and Localization Using MIMO Radars and Sonars},'' \emph{IEEE Transactions on Signal Processing}, vol.~54, no.~10, pp. 3873--3883, 2006.

\bibitem{4359542}
J.~Li, L.~Xu, P.~Stoica, K.~W. Forsythe, and D.~W. Bliss, ``{Range Compression and Waveform Optimization for MIMO Radar: A CramÉr–Rao Bound Based Study},'' \emph{IEEE Transactions on Signal Processing}, vol.~56, no.~1, pp. 218--232, 2008.

\bibitem{boyd2004convex}
S.~Boyd and L.~Vandenberghe, \emph{{Convex Optimization}}.\hskip 1em plus 0.5em minus 0.4em\relax Cambridge university press, 2004.

\bibitem{10050406}
Z.~Wang, X.~Mu, and Y.~Liu, ``{STARS Enabled Integrated Sensing and Communications},'' \emph{IEEE Transactions on Wireless Communications}, vol.~22, no.~10, pp. 6750--6765, 2023.

\bibitem{9124713}
X.~Liu, T.~Huang, N.~Shlezinger, Y.~Liu, J.~Zhou, and Y.~C. Eldar, ``{Joint Transmit Beamforming for Multiuser MIMO Communications and MIMO Radar},'' \emph{IEEE Transactions on Signal Processing}, vol.~68, pp. 3929--3944, 2020.

\bibitem{8288677}
F.~Liu, C.~Masouros, A.~Li, H.~Sun, and L.~Hanzo, ``{MU-MIMO Communications With MIMO Radar: From Co-Existence to Joint Transmission},'' \emph{IEEE Transactions on Wireless Communications}, vol.~17, no.~4, pp. 2755--2770, 2018.

\bibitem{boumal2023intromanifolds}
N.~Boumal, \emph{{An introduction to optimization on smooth manifolds}}.\hskip 1em plus 0.5em minus 0.4em\relax Cambridge University Press, 2023.

\bibitem{10.5555/1557548}
P.-A. Absil, R.~Mahony, and R.~Sepulchre, \emph{{Optimization Algorithms on Matrix Manifolds}}.\hskip 1em plus 0.5em minus 0.4em\relax USA: Princeton University Press, 2007.

\bibitem{GVK502988711}
J.~Nocedal and S.~Wright, \emph{{Numerical optimization}}, 2nd~ed., ser. Springer series in operations research and financial engineering.\hskip 1em plus 0.5em minus 0.4em\relax New York, NY: Springer, 2006.

\bibitem{Sato03042015}
H.~Sato and T.~Iwai, ``{A new, globally convergent Riemannian conjugate gradient method},'' \emph{Optimization}, vol.~64, no.~4, pp. 1011--1031, 2015.

\bibitem{sun2006short}
D.~Sun, ``{A Short Summer School Course on Modern Optimization Theory: Optimality Conditions and Pertubation Analysis, Part I, Part II, Part III},'' Lecture Notes, National University of Singapore, Singapore, 2006.

\bibitem{1406483}
S.~Pillai, T.~Suel, and S.~Cha, ``{The Perron-Frobenius theorem: some of its applications},'' \emph{IEEE Signal Processing Magazine}, vol.~22, no.~2, pp. 62--75, 2005.

\bibitem{7031971}
E.~Björnson, L.~Sanguinetti, J.~Hoydis, and M.~Debbah, ``{Optimal Design of Energy-Efficient Multi-User MIMO Systems: Is Massive MIMO the Answer?}'' \emph{IEEE Transactions on Wireless Communications}, vol.~14, no.~6, pp. 3059--3075, 2015.

\bibitem{9667503}
N.~Zhao, Y.~Wang, Z.~Zhang, Q.~Chang, and Y.~Shen, ``{Joint Transmit and Receive Beamforming Design for Integrated Sensing and Communication},'' \emph{IEEE Communications Letters}, vol.~26, no.~3, pp. 662--666, 2022.

\bibitem{9805471}
Z.~Huang, K.~Wang, A.~Liu, Y.~Cai, R.~Du, and T.~X. Han, ``{Joint Pilot Optimization, Target Detection and Channel Estimation for Integrated Sensing and Communication Systems},'' \emph{IEEE Transactions on Wireless Communications}, vol.~21, no.~12, pp. 10\,351--10\,365, 2022.

\bibitem{10138058}
X.~Song, J.~Xu, F.~Liu, T.~X. Han, and Y.~C. Eldar, ``{Intelligent Reflecting Surface Enabled Sensing: Cramér-Rao Bound Optimization},'' \emph{IEEE Transactions on Signal Processing}, vol.~71, pp. 2011--2026, 2023.

\bibitem{1143830}
R.~Schmidt, ``{Multiple emitter location and signal parameter estimation},'' \emph{IEEE Transactions on Antennas and Propagation}, vol.~34, no.~3, pp. 276--280, 1986.

\bibitem{10135096}
Z.~Wang, X.~Mu, and Y.~Liu, ``{Near-Field Integrated Sensing and Communications},'' \emph{IEEE Communications Letters}, vol.~27, no.~8, pp. 2048--2052, 2023.

\bibitem{10251151}
Z.~Ren, Y.~Peng, X.~Song, Y.~Fang, L.~Qiu, L.~Liu, D.~W.~K. Ng, and J.~Xu, ``{Fundamental CRB-Rate Tradeoff in Multi-Antenna ISAC Systems With Information Multicasting and Multi-Target Sensing},'' \emph{IEEE Transactions on Wireless Communications}, vol.~23, no.~4, pp. 3870--3885, 2024.

\end{thebibliography}
\end{document}